\documentclass[12pt, onecolumn, journal]{IEEEtran}
\usepackage{amsmath,amsfonts}
\usepackage{amssymb}
\usepackage{algorithm}
\usepackage{array}
\usepackage[caption=false,font=normalsize,labelfont=sf,textfont=sf]{subfig}
\usepackage{textcomp}
\usepackage{stfloats}
\usepackage{siunitx}
\usepackage{url}
\usepackage{verbatim}
\usepackage{graphicx}
\usepackage{cite}

\usepackage{comment}

\usepackage{wrapfig}
\usepackage{bm}
\usepackage{amsmath}
\usepackage{algpseudocode}

\usepackage[acronym,,nohypertypes={acronym,notation}]{glossaries}
\newacronym{3d}{3-D}{three-dimensional}
\newacronym{3gpp}{3GPP}{3rd Generation Partnership Project}
\newacronym{5g}{5G}{fifth generation}
\newacronym{6g}{6G}{sixth generation}
\newacronym{aoa}{AoA}{angle of arrival}
\newacronym{aod}{AoD}{angle of departure}
\newacronym{as}{AS}{antenna selection}
\newacronym{awgn}{AWGN}{additive white gaussian noise}
\newacronym{af}{AF}{Amplify-and-Forward}
\newacronym{bs}{BS}{base station}
\newacronym{bcd}{BCD}{block coordinate descent}
\newacronym{b5g}{B5G}{Beyond-5G}
\newacronym{csi}{CSI}{channel state information}
\newacronym{chest}{CHEST}{channel estimation}
\newacronym{cdf}{CDF}{cumulative density function}
\newacronym{dl}{DL}{downlink}
\newacronym{da}{D/A}{Diginal/Analog}
\newacronym{df}{DF}{Decode-and-Forward}
\newacronym{ee}{EE}{energy efficiency}
\newacronym{er}{ER}{ergodic rate}
\newacronym{eh}{EH}{enerhy harvesting}
\newacronym{epa}{EPA}{equal power allocation}
\newacronym{fp}{FP}{fractional programming}
\newacronym[plural=FETs, firstplural=field-effect transistor (FETs)]{fet}{FET}{ field-effect transistor}
\newacronym{fpga}{FPGA}{field programmable gate array}
\newacronym{flops}{FLOPs}{Floating Point Operations }
\newacronym{ga}{GA}{genetic algorithm}
\newacronym{iid}{i.i.d.}{independent and identically distributed}
\newacronym{isac}{ISAC}{integrated sensing and communication}
\newacronym{iot}{IoT}{Internet of Things}
\newacronym{kkt}{KKT}{Karush–Kuhn–Tucker}
\newacronym{los}{LoS}{line-of-sight}
\newacronym{ldt}{LDT}{lagrangian dual transform}
\newacronym{mimo}{MIMO}{multiple-input multiple-output}
\newacronym{m-mimo}{M-MIMO}{massive MIMO}
\newacronym{mmwave}{mmWave}{millimeter-wave}
\newacronym{mcs}{MCs}{Monte-Carlo Simulation}
\newacronym{miso}{MISO}{Multiple-input single-output}
\newacronym{mm}{MM}{majorization-minimization}
\newacronym{mr}{MR}{maximum ratio}
\newacronym{pc}{PC}{pilot contamination}
\newacronym{pdd}{PDD}{penalty dual decomposition}
\newacronym{pso}{PSO}{particle swarm optimization}
\newacronym{ps}{PS}{power splitting}
\newacronym{qos}{QoS}{Quality of Service}
\newacronym{rf}{RF}{radio frequency}
\newacronym{re}{RE}{resource efficiency}
\newacronym[plural=RISs]{ris}{RIS}{reconfigurable intelligent surface}
\newacronym{se}{SE}{spectral efficiency}
\newacronym{sinr}{SINR}{signal-to-interference-plus-noise ratio}
\newacronym{snr}{SNR}{signal-to-noise ratio}
\newacronym{sca}{SCA}{sucessive convex approximation}
\newacronym{socp}{SOCP}{second order conic programming}
\newacronym{sdr}{SDR}{semi-defined relaxation}
\newacronym{sfp}{SFP}{sequential fractional programming}
\newacronym{sdp}{SDP}{semi-definite program}
\newacronym{sr}{SR}{sum-rate}
\newacronym{tdd}{TDD}{time-division duplex}
\newacronym[plural=UEs, firstplural=users' equipment (UEs)]{ue}{UE}{user's equipment}
\newacronym{ul}{UL}{uplink}
\newacronym{upa}{UPA}{uniform planar array}
\newacronym{uspa}{USPA}{uniform squared planar array}
\newacronym{ula}{ULA}{uniform linear array}
\newacronym{vr}{VR}{visibility region}
\newacronym{wsr}{WSR}{weigthed sum-rate}
\newacronym{xl-mimo}{XL-MIMO}{extra-large scale massive MIMO}
\newacronym{xr}{XR}{extended reality }
\newacronym{zf}{ZF}{zero-forcing}

\usepackage{array}
\usepackage{tabularx}
\usepackage{multirow}
\usepackage{tikz}
\definecolor{gold}{rgb}{0.85,.66,0}
\definecolor{cian}{rgb}{.02,.7,.95}
\definecolor{ppp}{rgb}{.7,.3,.82}

\begin{document}

\title{Energy-Efficient Active Element Selection in RIS-aided Massive MIMO Systems}
\author{Wilson Souza Jr,  {Taufik Abrão}, José Carlos Marinello
\thanks{J. C. Marinello is with the Electrical Engineering Department, Federal University of Technology PR, Cornélio Procópio, PR, Brazil.  (e-mail: faugustobueno@gmail.com, omluizalberto@gmail.com, jcmarinello@utfpr.edu.br).}
\thanks{Wilson Souza Jr and T. Abrao are with the Department of Electrical Engineering Londrina State University, Parana Brazil (e-mail  taufik@uel.br)}}

\maketitle

\begin{abstract}
This paper delves into the critical aspects of optimizing \gls{ee} in active \gls{ris}-assisted \gls{m-mimo} wireless communication systems. We develop a comprehensive and unified theoretical framework to analyze the boundaries of \gls{ee} within \gls{m-mimo} systems integrated with active \gls{ris} while adhering to practical constraints. Our research focuses on a formulated \gls{ee} optimization problem aiming to maximize the \gls{ee} for active \gls{ris}-assisted \gls{m-mimo} communication systems. Our goal is strategically finding the number of active \gls{ris} elements for outperforming the \gls{ee} attainable by an entirely passive \gls{ris}. Besides, the proposed novel solution has been specifically tailored to the innovative problem considered. The formulation and solution design take into account
analytical optimization techniques, such as \gls{ldt} and \gls{fp} optimization, facilitating the effective implementation of RIS-aided \gls{m-mimo} applications in real-world settings. In particular, our results show that the proposed algorithm is able to provide up to 120$\%$ higher \gls{ee} in comparison to the entirely passive \gls{ris}. Besides, we found that the active \gls{ris} can operate with lower than half of reflecting elements with respect to the entirely passive \gls{ris}. Finally, in view of active \gls{ris} achieving the complete utilization of amplification power available, it should be equipped with a reasonable number of reflecting elements, above $N=49$.
\end{abstract}


\section{Introduction}\label{sec:intro}
The proliferation of the \gls{5g} wireless network has catalyzed worldwide incentive and development efforts in research, resulting in an intensified exploration of possibilities offered by the \gls{6g} wireless network. Projected to be characterized and recognized by exceptional data-driven capabilities and instantaneous connectivity global ubiquity, \gls{6g} is expected to provide a wide range of applications and services that will revolutionize communication systems. These innovations span across diverse domains, such as \gls{xr}, holographic communications, etc., promising to redefine the technological scenario. The advent of \gls{6g} will bring forth plenty of new services and withal elevate existing ones to higher levels. However, in view of reaching this aim, the integration of cutting-edge technologies becomes essential to facilitate the deployment of these services and ensure their optimal functioning in the technological scenario. In essence, the commercialization of \gls{6g} will necessitate a proactive approach in adopting state-of-the-art technologies, marking a paradigm shift in how we envision wireless communication networks' potential.

Among all emerging technologies, \gls{ris} stands out as one of the most promising recent enabling techniques poised to revolutionize future communication systems, extending beyond the \gls{5g} and \gls{6g} eras. \gls{ris} has the potential to significantly enhance transmission capabilities by establishing a ``virtual" re-configurable/re-programmable communication link. Essentially the \gls{ris}, is a thin planar surface consisting of a massive number of low-cost reflecting elements composed of meta-materials, to reconfigure and redirect the impinging signal, by changing its electromagnetic properties (phase and/or magnitude) \cite{8910627}, so that, the reflected signal can be focused toward a specific direction with a specific gain. The electromagnetic properties of the impinging waves are precisely controlled by utilizing electronic components such as PIN diodes or \gls{rf} switches integrated into the \gls{ris} panel. For instance, changing the impedance of PIN diodes or tuning the \gls{rf} switch facilitates the manipulation of the reflection with respect to its direction and amplitude, respectively. This dynamic adjustment serves to enhance specific target metrics. Therefore, \gls{ris} is promising due to its capacity to control in a software-defined manner the wireless environment and its potential to mitigate certain intrinsic effects. The specific advantages of the \gls{ris} arrays can be summarized as follows \cite{9122596}: \\
1) Easiness in implementation; \\
2) \gls{se} and \gls{ee} enhancement; \\
3) Sustainability; \\
4) Compatibility.

Due to the aforementioned characteristics, \gls{ris}s are recognized as a potential solution for solving a wide range of challenges in commercial and civilian applications. Besides, in contrast to conventional relaying methods such as \gls{af} and \gls{df}, \gls{ris} presents many advantages. It offers a cost-effective yet efficient solution that optimizes both system's \gls{ee} and \gls{se}. On this matter, like any emerging technology, new applications and a range of problems arise as a result. Besides, from the perspective of communication and signal processing, open problems such as phase-shift optimization of scattering elements, and channel acquisition of additional links imposed by deploying the \gls{ris} are extremely necessary to be addressed and efficiently solved.

Although \gls{ris} presents a range of advantages, there are some critical issues that are found in real-world applications. Consequently, various architectural variants for the \gls{ris} have been introduced to address these challenges and enhance its overall effectiveness. These variations aim to significantly enhance performance, offering innovative solutions to overcome challenges and elevate the effectiveness of \gls{ris}-supported communication systems. Among them, we can remark the active \gls{ris} structure, which is an alternative for providing better conditions for the double-fading attenuation link. In active \gls{ris}, each reflecting element is supported by a set of active-load impedances that enable the active \gls{ris} to operate as an active reflector reflecting and applying power amplification on the impinging signal. An interesting advantage of the active \gls{ris} elements is their capability to estimate the channel between the transmitter-\gls{ris} in \gls{dl} and \gls{ul} at the \gls{ris} side \cite{9370097}. Besides, since the active \gls{ris} has the ability to amplify the incident signal, fewer reflecting elements are required to achieve the same \gls{snr} at the receiver when compared with the conventional entirely passive \gls{ris} \cite{9377648}. However, evaluating the \gls{ee} for the entirely passive \gls{ris} versus active \gls{ris} is an important problem that should be further investigated. 

Conversely, the proliferation of \gls{m-mimo} technology, characterized by the deployment of an extensive array of transmit/receiver antennas at \gls{bs}, plays a pivotal role in enhancing the \gls{se} of the system. This technology is instrumental in efficiently accommodating multiples \glspl{ue} by leveraging the same physical resources. This strategic utilization enhances the \gls{se} and optimizes the resource utilization, exemplifying the versatility and efficiency of \gls{m-mimo} in contemporary wireless communication systems.

Besides the aforementioned technologies, a promising technology to be present in future communications networks is the \gls{mmwave} technology.  It emerges as a strategic pivotal solution in addressing the evolving connectivity demands. Operating within the high-frequency spectrum, typically spanning from 30 to 100 GHz, \gls{mmwave} technology presents a promising solution capable of supplying characteristics such as the utilization of higher bandwidth for contemporary communication networks. Furthermore, with the large arrays, the near-field emerges, enabling the beam focusing instead of the traditional beam steering for far-field communications \cite{an2023beamfocusingaided,9921216,10123941}.

However, the implementation of these higher-frequency systems, which often entails the use of large-scale arrays at the \gls{bs} or the \gls{ris}, introduces inherent complexities in both passive and active beamforming, demanding low-complexity, approximate yet efficient solutions for accurate channel estimation and \gls{ris} configuration. 
The joint operation of \gls{ris}, \gls{m-mimo}, and \gls{mmwave} technologies emerges as key enablers for future communication systems. Expected to play central roles in the upcoming generation of wireless networks, these technologies are designed to support high data rates and accommodate a substantial number of connected \glspl{ue}, operating with expanded array sizes. In response to this evolution, there is a fundamental growth to prioritize the system's \gls{ee}, measured in bits-per-Joule\footnote{Alternatively, Joule-per-bit.}, as a crucial performance metric.
The focus on \gls{ee} emphasizes the matter of developing sustainable communication systems, especially as efficient energy utilization becomes a global concern across various fields, including telecommunications. In function of alarming levels of climate change, there is an urgent need to limit global warming to 1.5 degrees above the pre-industrial mark \cite{9467318}. Within this context, the development of the next-generation telecommunication standard, often referred to as \gls{b5g} or \gls{6g}, assumes particular importance. It aims to establish a highly energy-efficient system capable of meeting expected requirements, such as a peak throughput of 1 [Tbps] and latency in the order of microseconds \cite{8922617}. This forward-looking approach aligns with the broader goal of enabling technological advancements that not only meet the requirements necessary in terms of performance but actively contribute to the sustainability challenges of our time.

In light of these challenges, the development of pioneering strategies that not only enhance \gls{ee} but also facilitate the intricate processes of channel estimation and configuration for large-scale array systems is of paramount importance.  In this paper, our focus stays dedicated to addressing these challenges with innovative approaches and practical solutions tailored to the modern network's infrastructures.

\subsection{Contributions of the paper}
The contributions of this  paper lie in analyzing the \gls{ris}-assisted \gls{m-mimo} scenario, which will become essential components of future communication systems. Specifically, the aim is to devise a solution that effectively optimizes the signal amplification of active \gls{ris} concerning the \gls{ee}, envisaging the determination of the number of active \gls{ris} elements that outperforms entirely passive \gls{ris}. Elaborate further, in this paper, we systematically:

\begin{itemize}

\item Propose a comprehensive optimization methodology for system \gls{ee} maximization, focusing on the deployment of specific optimization tools, including convex optimization procedures in view of reducing the computational complexity aiming the application of the algorithm in a real-world scenario;

\item Optimize the phase and amplitude of the active \gls{ris} in a sustainable perspective, $i.e.$, the phase and amplitude that results in the power consumption necessary to obtain high-performance gains sufficient to outperform the totally passive \gls{ris};

\item Investigate the number of \gls{ris} active elements to be selected that surpasses the \gls{ee} performance of the entirely passive \gls{ris} within the system \gls{ee} optimization framework and different configurations/scenarios parameters.

\end{itemize}

\subsection{Organization}
The remainder of this  paper is organized into seven sections.  These sections are meticulously structured to provide a comprehensive exploration of the various aspects of the \gls{ris}-assisted \gls{m-mimo} system. In {\bf Section \ref{sec:relatedWorks}}, we summarize the related works; while in {\bf Section \ref{sec:gen_sys_model}}, we present the groundwork by formulating a general system model that effectively captures the key aspects associated with the integration of the active \gls{ris} into the \gls{m-mimo} systems. Also, we explore the \gls{se} and \gls{ee} definitions and present the coupling power consumption model for \gls{ris}-assisted \gls{m-mimo} systems. In {\bf Section \ref{sec:optzTech}}, we review the definitions, features, and concepts of some fundamental techniques for addressing our intended problem, such as \gls{fp} techniques and sum-of-ratio techniques. Moving forward, in {\bf Section \ref{sec:ProblemFormulation}}, we meticulously present the defined problem formulation, emphasizing its relevance and significance in contemporary real-world applications. The motivations and justifications for adopting this particular problem are thoroughly expounded, shedding light on the specific challenges and opportunities that emphasize its importance in modern communication systems.
{\bf Section \ref{sec:Solution}} provides a complete and detailed description of our proposed solution for the \gls{ee} problem in active \gls{ris}-assisted \gls{m-mimo} systems.
{\bf Section \ref{sec:Result}} presents numerical results, accompanied by an extensive and in-depth discussion on the effectiveness, efficiency, and potential of the proposed technique.  The section also includes a relevant discussion on the advantages/drawbacks of the active \gls{ris} structure in contrast to the entirely passive \gls{ris} within the context of \gls{ee}.

Finally, in {\bf Section \ref{sec:concl}}, we provide a relevant and comprehensive discussion that encapsulates our perspective into the problem-solving approach adopted throughout this paper as well as the obtained insights. Our analysis and conclusions are framed within the broader context of the current state of the art, aiming to provide valuable insights and potential directions for future research.

\section{Related Works}\label{sec:relatedWorks}

An extensive number of works including \cite{8982186,9090356,Zhi2022,9369326,10371348,10183034,He2022,9814527,9802804,10136805,9246254,9681803, 10008751, liu2024joint, 9810984, 9998527, ye2023energy, 9963962} have studied the \gls{ris} deployment impact in \gls{m-mimo} systems under different objectives, such as: maximizing the \gls{wsr} \cite{8982186,9090356}, maximizing the \gls{sr} \cite{Zhi2022}, maximizing the \gls{ee} \cite{9369326, 10371348,10183034}, \gls{ris}-assisted sensing \cite{He2022,9814527}, minimizing the total transmit power \cite{9802804}, maximizing the minimum rate \cite{10136805,9246254}, for interference nulling \cite{9681803,10008751}, and active \gls{ris} \cite{liu2024joint, 9810984, 9998527, ye2023energy, 9963962}. 

The above-cited studies demonstrate the immense potential and versatility of \gls{ris} within wireless systems.  The work in \cite{8982186} investigates the weighted \gls{sr} maximization. It uses a convex optimization approach to design the \gls{bs} precoding matrix and the \gls{ris}-phase shift for the passive \gls{ris}. The paper proposes two different techniques for the analyzed problem.  The proposed optimization method utilizes \gls{fp} techniques aiming at the problem "convexation". Besides, an \gls{bcd} method is utilized to optimize sequentially the \gls{bs} precoding matrix and the \gls{ris} phase shift, where analytical methods and {\it Riemannian} conjugate gradient have been utilized. The proposed method is available under two different scenarios: perfect and imperfect channel estimation.  Numerical simulations confirm the effectiveness and efficiency of the proposed method.

The work in \cite{9090356} suggests that by attempting to reduce inter-cell interference, the weighted \gls{sr} in \gls{ris}-aided \gls{dl} multicell could be maximized. Hence, to achieve this aim, \cite{9090356} jointly
optimizes the active precoding matrices at \gls{bs}s and the phase shifts at the \gls{ris} subject to each \gls{bs}’s power constraint and unit modulus constraint, respectively. In this case, the optimization problem does
not consider any \gls{qos} restriction, $e.g.$, a minimum rate requirement for the users. As a result, this solution tends to maximize the rate of users with better channel conditions.

In \cite{Zhi2022}, the authors investigated the ergodic rate of an \gls{ris}-assisted \gls{m-mimo} system. Applying random matrix theory, the authors derive a closed-form lower bound expression for the system ergodic rate. The derived bound reveals tightness over several system and channel parameters. Besides, it is promising since the statistical \gls{csi} optimization can be deployed based on this bound, being useful to alleviate the channel estimation burden since it matches very well with the expected results. Besides, they provided a gradient-ascent method to optimize the \gls{ris} angle phase shift in view of maximizing the system \gls{sr}.

The \gls{ee} maximization problem is addressed in \cite{9369326} by considering \gls{ul} multi-user single-antenna devices and \gls{miso} \gls{ris}-aided systems. The joint optimization involves the transmit power at users, phase shift at the \gls{ris}, and combining matrix at \gls{bs} receiver. The proposed \gls{bcd} iterative procedure can handle one variable during each iteration while the others are held fixed. Numerical results indicate that the proposed scheme outperforms baselines that jointly optimize two of three variables.

In \cite{10371348}, the joint number of antennas, power allocation, and passive \gls{ris} reflecting coefficients optimization for \gls{ee} maximization with statistical \gls{csi} is studied. Analytical techniques have been applied in view of maximizing the \gls{ee} with low overhead for channel estimation. \gls{ee} problem has been maximized with a low-complexity algorithm, where the power allocation, \gls{ris} phase shift, and the number of active antennas at \gls{bs} have been optimized based on closed-form expressions, utilizing \gls{bcd} method. The method was demonstrated to be promising, outperforming the gradient ascend methodology for passive \gls{ris} reflecting coefficients optimization with statistical \gls{csi}.

In \cite{10183034}, the authors also analyzed the \gls{ee} in \gls{ris}-assisted systems. In this study, the \gls{ee} maximization has been analyzed under joint optimization of power consumption for both \gls{ris} and \gls{bs}. The authors proposed an innovative power consumption model on the \gls{ris} by considering the ON and OFF state of \gls{ris} unit reflecting element, matching further precisely with the real-world behavior of physical components. The paper also proposes low-complexity and effective algorithms, which have been proven by the simulation results to be promising. 

In \cite{He2022}, the authors studied the  \gls{isac} system for a double \gls{ris}-assisted system. The research primarily focuses on addressing the challenge of mutual interference between the radar and communication functionalities by leveraging a joint optimization approach for beamforming and radar operations. To manage the complexity of the solution, the authors introduce a \gls {pdd}-based approach and a \gls{bcd} algorithm. These methods effectively reduce computational complexity, offering practical and efficient solutions for real-world implementation. Moreover, the work presented in \cite{9814527} addresses the challenges of a sensing-assisted multi-user \gls{mmwave} system. The study focuses on optimizing the channel sensing duration for each \gls{ue} to enhance the overall system \gls{sr}. By effectively managing the sensing duration, the research aims to improve the performance of the \gls{mmwave} system, offering potential advancements for future wireless communication technologies.

In \cite{9802804}, the phase shift and \gls{bs} beamforming optimization techniques have been proposed in view of minimizing the \gls{ul} transmit power in an \gls{ris}-aided \gls{iot} network. The proposed technique was demonstrated to be promising since it can reduce to about half of the \gls{ul} transmit power over the conventional scheme without \gls{ris}. The methodology exploits the product {\it Riemannian manifold} structure of the sets of unit-modulus phase shifts and unit-norm beamforming vectors. Besides, the {\it manifold} method converts the non-convex \gls{ul} transmit power minimization problem into an unconstrained problem and then finds the optimal solution over the product {\it Riemannian manifold}.

In \cite{10136805}, the focus was on the study of the maximum-minimum rate problem. The research delved into the joint optimization of transmit and passive beamforming, considering the uncertainty of \gls{ue}s locations. To address this challenge, the authors proposed an alternating optimization approach combined with \gls{sca} technique and a penalty algorithm employing a double loop.

In the research paper \cite{9246254}, the authors demonstrated that the non-convex $\mathrm{max}-\mathrm{min}$ rate problem in \gls{ris}-assisted multi-user communication systems can be reformulated into a \gls{socp} problem. They also explored the application of \gls{sdr} and \gls{sca} techniques to optimize the passive and active beamforming for the system under consideration.

In \cite{9681803}, the \gls{ris} structure is deployed aiming at interference nulling, where an alternating projection method is utilized. This method is able to converge for a solution that completely eliminates the interference. However, the number of elements for achieving this goal can vary according to some parameters, such as the number of served \glspl{ue}. Besides, the authors also explored the \texttt{max-min} rate problem. The numerical results are revealed to be promising since the proposed procedure achieves high performance with low complexity. 
Adhering to a similar approach, the authors in \cite{10008751} proposed a method to determine the angles for two distinct passive \glspl{ris}, aiming to achieve complete orthogonalization of channels between \glspl{ue} and thereby nulling interference. The investigation delves specifically into both diagonal and beyond-diagonal \glspl{ris}, employing a {\it Manifold}-based algorithm to find the angles for both \glspl{ris}, adhering to the unitary matrix constraint essential for achieving complete orthogonalization of channels between the \glspl{ue}. Notably, the study highlights that signal amplification is unnecessary for achieving orthogonalization. Furthermore, the paper introduces an efficient channel estimation method to complement the proposed approach.

Our previous related works and contributions on the \gls{re} optimization, \gls{ee} and \gls{se} in \gls{m-mimo} systems aided or not by \gls{ris} include: \cite{10371348, 10040750, DOSSANTOS2022101646, 9718019, 10070317, 9187980, ligia, ubiali2021energy, 10436034, rosa2023improving}.  In this paper, we propose to deploy the \gls{fp} methodology to solve the phase-shift and amplification coefficients of an active \gls{ris} aiming to \gls{ee} maximization problem considering the context of the \gls{dl} \gls{ris}-aided multiuser \gls{m-mimo} systems.

\section{General System Model for RIS-aided M-MIMO}\label{sec:gen_sys_model}

We start the section by describing the general model for \gls{m-mimo} systems assisted by an \gls{ris} setup considered. This system model is conceived to cover a diverse range of practical scenarios. Subsequently, we present the models for the \gls{dl} received baseband signals. Built such model, we finish the section by formulating expressions for \gls{se} and \gls{ee} within the specified \gls{ris}-aided \gls{m-mimo} scenarios. The notation deployed in this paper is entirely outlined in Table \ref{tab:system-model-symbols}.

\begin{table}[!htbp]
\caption{List of notation and symbols for the \gls{ris}-aided \gls{m-mimo} Scenario
\label{tab:system-model-symbols}}{%
\begin{tabular}{@{}|rl|@{}}
\hline
\textbf{Symbol} & \textbf{Description}\\
\hline
$M \in \mathbb{Z}_+$ & Number of \gls{bs} antennas\\
$K \in \mathbb{Z}_+$ & Number of \glspl{ue}\\
$N \in \mathbb{Z}_+$ & Number of total \gls{ris} elements\\
$\mathcal{K} = \{1,2,\dots,K\}$ & \glspl{ue} set\\
$\mathcal{N} = \{1,2,\dots,N\}$ & \glspl{ris} elements set\\
$\lambda > 0$ & Carrier wavelength\\
$\beta_{1} \in \mathbb{R}$ & Large-scale fading coefficient of the channel from \gls{bs} to \gls{ris}\\
$\beta_{2,k} \in \mathbb{R}$ & Large-scale fading coefficient of the channel from \gls{ris} to $k$-th \gls{ue}\\
$\boldsymbol{H} \in \mathbb{C}^{M\times K}$ & Cascaded channel matrix of the \gls{bs} and \glspl{ue} through \gls{ris}\\
$\boldsymbol{H}_1 \in \mathbb{C}^{M\times N}$ & Channel between the \gls{bs} and \gls{ris} \\
$\bar{\boldsymbol{H}}_1 \in \mathbb{C}^{M\times N}$ & \Gls{los} component of the \gls{bs} and \gls{ris} channel \\
$\tilde{\boldsymbol{H}}_1 \in \mathbb{C}^{M\times N}$ & Multipath component of the \gls{bs} and \gls{ris} channel \\
$\boldsymbol{H}_{2} \in \mathbb{C}^{N\times K}$ & Channel between the \gls{ris} and \gls{ue}s \\
$\bar{\boldsymbol{H}}_{2} \in \mathbb{C}^{N\times K}$ & \Gls{los} component of the \gls{ris} and \gls{ue}s channel \\
$\tilde{\boldsymbol{H}}_{2} \in \mathbb{C}^{N\times K}$ & Multipath component of the \gls{ris} and \gls{ue}s channel \\
$\boldsymbol{v} \in \mathbb{C}^{N}$ & \gls{ris} amplitude/phase-shift vector \\
$\boldsymbol{\Phi} \in \mathbb{C}^{N \times N} $ & \gls{ris} amplitude/phase-shift diagonal matrix  \\
$\kappa_1 \in [0, \infty)$ & Rician $\kappa$-factor for \gls{ris} and \gls{bs} channel \\
$\kappa_{2,k} \in [0, \infty)$ & Rician $\kappa$-factor for \gls{ris} and $k$-th \gls{ue}  \\
$p_k$ $\in$ $\mathbb{R}^+$ & Power designated to the $k$-th \gls{ue}\\
$\boldsymbol{W}$ $\in$ $\mathbb{C}^{M\times K}$ & Precoding matrix \\
$s_k$ $\in$ $\mathbb{C}$ & Information symbol intended to the $k$-th \gls{ue}\\
$n_k$ $\in$ $\mathbb{C}$ & \gls{awgn} sample at the $k$-th \gls{ue}\\
$\boldsymbol{z}$ $\in$ $\mathbb{C}^{N \times 1}$ & \gls{awgn} sample vector at the \gls{ris}\\
$\alpha_n$ $\in \mathbb{R}_{+}$ & Amplitude value of $n$-th element of \gls{ris}\\
$\theta_n$ $\in [0,2\pi)$ & Phase-shift value of $n$-th element of \gls{ris}\\
$\alpha_{\max}$ $\in \mathbb{R}_{+}^{*}$ & Maximum amplitude provided by the \gls{ris}\\
$P_{\max}^{\rm RIS}$ $\in \mathbb{R}_{+}$ & Maximum amplification power of \gls{ris}\\
\hline
\end{tabular}
}
\end{table}

Consider a \gls{dl} \gls{ris}-assisted \gls{m-mimo} scenario where the direct link between the \gls{bs} and \glspl{ue} is completely obstructed, thus, $K$ single antenna \glspl{ue} are served by the \gls{bs} with $M$ elements through an \gls{ris} with $N$ elements.
 Besides, we assume a direct connection between the \gls{ris} and the \gls{bs} is made by means of a dedicated link via a programmable controller, enabling the \gls{bs} to manage both phase shifts and amplitude coefficients of the \gls{ris}. Furthermore, we assume that the time necessary for \gls{ris} amplitude/phase configuration is shorter than the channel {\it coherence time}, $i.e.$, at the start of each time slot, the \gls{bs} should configure the \gls{ris} appropriately for its operation.  Let us denote $\boldsymbol{H}_1$ as the channel from the \gls{bs} to the \gls{ris} and $\boldsymbol{H}_2$ as the channel from the \gls{ris} to the \glspl{ue}. The cascaded channel between the \gls{bs} and \glspl{ue} through the \gls{ris} can be given as $\boldsymbol{H} = \left[\boldsymbol{h}_1, \boldsymbol{h}_2, \dots, \boldsymbol{h}_K \right]$; more specifically, denoted as
\begin{equation}
     \boldsymbol{h}_k = \boldsymbol{H}_1 \boldsymbol{\Phi} \boldsymbol{h}_{2,k}, \quad \forall k \in \mathcal{K},
\end{equation}
where $\boldsymbol{\Phi} = {\rm diag}(\boldsymbol{v})$ $\in$ $\mathbb{C}^{N\times N}$, with $\boldsymbol{v} = [v_1^*, v_2^*, \dots, v_N^*]^H$ $\in$ $\mathbb{C}^{N \times 1}$, being $v_n = \alpha_n e^{j\theta_n}$ $\in$ $\mathbb{C}$ the phase-shift and amplitude applied by the $n$-th element of the \gls{ris}, with $\alpha_n$ $\in [0,\alpha_{\max}]$ and $\theta_n$ $\in [0,2\pi)$.
The \gls{bs} uses a precoding matrix $\boldsymbol{W} = \left[\boldsymbol{w}_1, \boldsymbol{w}_2, \dots, \boldsymbol{w}_K \right]$ $\in$ $\mathbb{C}^{M\times K}$, with $||\boldsymbol{w}_k||^2 = p_{k}$ $\in$ $\mathbb{R}^{+}$, in view of beamforming the signal towards the \glspl{ue}, where $p_k$ is the power allocated for the $k$-th \gls{ue} served. Here, for simplicity, we adopt \gls{mr} precoding with \gls{epa}.

Utilizing the system model explained above, the \gls{dl} baseband transmitted signal $\boldsymbol{x}$ $\in \mathbb{C}^{M\times 1}$ can be written as
\begin{equation}
    \boldsymbol{x} = \sum_{k=1}^K \boldsymbol{w}_k s_k, 
\end{equation}
 where $s_k$ is the symbol intended to the $k$-th \gls{ue}, with $\mathbb{E}[|s_k|^2] = 1$. While the received signal by the $k$-th \gls{ue} is given as
\begin{equation}
    y_k =  \boldsymbol{w}_k^H \boldsymbol{h}_k s_k  + \sum_{j=1, j\neq K }^{K} \boldsymbol{w}_j^H \boldsymbol{h}_k s_k +  \boldsymbol{h}_{2,k}^H \boldsymbol{\Phi} \boldsymbol{z} + n_k,
\end{equation}
in which $n_k$ $\in \mathbb{C}$ is the \gls{awgn} sample at the $k$-th \gls{ue}, following $n_k \sim \mathcal{CN}(0,\sigma^2)$, $\forall k \in \mathcal{K}$, and $\mathbf{z}$ $\in \mathbb{C}^{N\times 1}$ is the \gls{awgn} 
at the \gls{ris}, with $\mathbf{z} \sim \mathcal{CN}(\mathbf{0}_N,\sigma_{{\rm RIS}}^2\mathbf{I}_N)$ .

\subsection{RIS-aided M-MIMO Channel Modelling}

 The \gls{ris} takes the form of a thin \gls{uspa}, featuring $N = N_h \times N_v $ total elements, where $N_h = N_v = \sqrt{N}$ elements in the horizontal and vertical directions, with $\sqrt{N}$ being a positive integer ($\sqrt{N} \in \mathbb{Z}_+$). Furthermore, the elements are vertically and horizontally equally spaced with spacing $d_{\rm R}$. Besides, the \gls{bs} is assumed to be equipped with a \gls{ula} of $M$ total elements disposed in the $x$ axis, $i.e.$, parallel to the \gls{ris}. 

In the channel model for \gls{ris}-aided \gls{m-mimo}, we make the assumption of far-field wavefront propagation regime between the \gls{bs} and the \gls{ris} panels, as well as between the \gls{ris} and the \gls{ue}s, $i.e.$, the planar wavefront is assumed in both links. 

\vspace{2mm}
\noindent \textbf{BS-RIS link}: to be more specific, we establish the channel between the \gls{bs} and the \gls{ris} based on the Rician-fading model, given by \cite{Wu19, You20, Jiang23}:
\begin{equation}
    \boldsymbol{H}_1 = \sqrt{\frac{ \beta_{1} \kappa_1}{\kappa_1 + 1} } \bar{\boldsymbol{H}}_1 + \sqrt{\frac{\beta_{1}}{\kappa_1 + 1} } \tilde{\boldsymbol{H}}_1,
\end{equation}
where $\bar{\boldsymbol{H}}_1$ $\in \mathbb{C}^{M \times N}$ corresponds to the deterministic \gls{los} channel component, $\tilde{\boldsymbol{H}}_1$ $\in \mathbb{C}^{M \times N}$ to the Rayleigh-fading multipath component, $\kappa_1 \geq 0$ denotes the Rician $\kappa$-factor, and $\beta_{1}$ is the large-scale fading coefficient for this link specifically. The large-scale fading coefficient is given by
\begin{equation}
    \beta_1 = \frac{\beta_0}{d_{\rm BR}^{\lambda_{\rm BR}}},
\end{equation}
in which $\beta_0$ is the path loss at a reference distance of 1 meter, $d_{\rm BR}$, and $\lambda_{\rm BR}$ is the distance and the path-loss exponent between the \gls{bs} and \gls{ris}, respectively.
Particularly, the deterministic \gls{los} component can be given through the steering vector of a two-dimensional \gls{upa}. Let us consider the \gls{ris} centered in the origin of the Cartesian coordinates; therefore, each column of \gls{los} channel matrix $\bar{\boldsymbol{H}}_1$ is given as 
\begin{equation}
        \bar{\boldsymbol{H}}_{1} = \boldsymbol{a}_{\rm B}(\vartheta_{\rm B}) \boldsymbol{a}_{\rm R}(\vartheta_{\rm R}, \varphi_{\rm R})^H,
\end{equation}
where $\vartheta_{\rm B}$ is the azimuthal \gls{aod} of the signal departing from the \gls{bs} towards the \gls{ris}, and $\vartheta_{\rm R}$ and $\varphi_{\rm R}$ are the azimuthal and elevation \gls{aoa} at the \gls{ris} from the \gls{bs}. The array response vector of the \gls{bs} \gls{ula} and of the \gls{ris} \gls{upa}, are defined, respectively as
\begin{equation}
    \boldsymbol{a}_{\rm B}(\vartheta_{\rm B}) = \left[1,e^{j\pi\frac{2d_{\rm B}}{\lambda} \sin \vartheta_{\rm B}}, \dots, e^{j(M-1)\pi  \frac{2d_{\rm B}}{\lambda} \sin \vartheta_{\rm B}} \right]^H,
\end{equation}
\begin{IEEEeqnarray}{rCl}
    \boldsymbol{a}_{\rm R}(\vartheta_{\rm R}, \varphi_{\rm R}) &=& \left[1,e^{j\pi\frac{2d_{\rm R}}{\lambda} \sin \vartheta_{\rm R} \cos \varphi_{\rm R}}, \dots, e^{j(\sqrt{N}-1)\pi  \frac{2d_{\rm R}}{\lambda} \sin \vartheta_{\rm R} \cos \varphi_{\rm R}} \right]^H \\ && \hspace{2cm} \otimes \left[1, e^{j\pi \frac{2 d_{\rm R}}{\lambda} \sin \varphi_{\rm R}}, \dots, e^{j(\sqrt{N}-1)\pi \frac{2 d_{\rm R}}{\lambda} \sin \varphi_{\rm R}} \right]^H, \nonumber 
\end{IEEEeqnarray}
with $d_{\rm B}$ being the distance between the antenna elements at the \gls{bs}. Moreover, the multipath component follows an uncorrelated Rayleigh distribution such that $\mathbf{\tilde{h}}_{1,i} \sim \mathcal{CN} (\mathbf{0}_N, \mathbf{I}_N)$, $\forall i \in \{1,\dots,M\}$, with $\tilde{\mathbf{h}}_{1,i}$ being the $i$-th column of matrix $\tilde{\mathbf{H}}_1$.

\vspace{2mm}
\noindent \textbf{RIS-UE link}: the link between the \gls{ris} and \glspl{ue}, $\boldsymbol{H}_2 = [\boldsymbol{h}_{2,1} \, \boldsymbol{h}_{2,2} \, \dots \,\boldsymbol{h}_{2,K}]$, is also assumed to be Rician fading and given as
\begin{equation}
    \boldsymbol{h}_{2,k} = \sqrt{\frac{\beta_{2,k}{\kappa}_{2,k}}{{\kappa}_{2,k}+1}} \bar{\boldsymbol{h}}_{2,k} + \sqrt{\frac{\beta_{2,k}}{{\kappa}_{2,k} + 1}}\tilde{\boldsymbol{h}}_{{2,k}}, \quad \forall k \in \mathcal{K}
\end{equation}
where $\bar{\mathbf{h}}_{{2,k}}$ $\in \mathbb{C}^{N \times 1}$ corresponds to the deterministic \gls{los} channel component between the \gls{ris} and the $k$-th \gls{ue}, $\tilde{\boldsymbol{h}}_{2,k}$ $\in \mathbb{C}^{N\times 1}$ to the Rayleigh-fading multipath component between \gls{ris} and $k$-th \gls{ue}, and $\kappa_{2,k}$ denotes the Rician $\kappa$-factor for this same link, with $\kappa_{2,k} \geq 0$ $\forall k$ $\in \mathcal{K}$. Besides, $\beta_{2,k}$ corresponds  to the large-scale fading from the \gls{ris} to the $k$-th \gls{ue}, given as 
\begin{equation}
    \beta_{2,k} = \frac{\beta_0}{d_{{\rm RU},k}^{\lambda_{\rm RU}}}, 
\end{equation}
with $d_{{\rm RU},k}$ being the distance between the \gls{ris} and $k$-th \gls{ue}, and $\lambda_{\rm RU}$ the path loss exponent between the \gls{ris} and \glspl{ue}.

\vspace{2mm}
\noindent \textbf{\gls{los} Component}: similarly to the \gls{bs} and \gls{ris} channel model, the deterministic \gls{los} component $\bar{\boldsymbol{h}}_{2,k}$ can be given by the steering vector:  
\begin{IEEEeqnarray}{rCl}
\bar{\boldsymbol{h}}_{2,k} &=& 
\boldsymbol{a}_{{\rm R},k}(\vartheta_{{\rm R},k}, \varphi_{{\rm R},k}) 
    \\
    &=& \left[1,e^{j\pi\frac{2d_{\rm R}}{\lambda} \sin \vartheta_{\rm R} \cos \varphi_{\rm R}}, \dots, e^{j(\sqrt{N}-1)\pi  \frac{2d_{\rm R}}{\lambda} \sin \vartheta_{\rm R} \cos \varphi_{\rm R}} \right]^H \otimes \left[1, e^{j\pi \frac{2 d_{\rm R}}{\lambda} \sin \varphi_{\rm R}}, \dots, e^{j(\sqrt{N}-1)\pi \frac{2 d_{\rm R}}{\lambda} \sin \varphi_{\rm R}} \right]^H. \nonumber
\end{IEEEeqnarray}

\subsection{\gls{se} and \gls{ee} Formulation in \gls{ris}-Assisted \gls{m-mimo}}
In this subsection, we introduce two critical and crucial performance metrics for analyzing the performance of communication systems. We initiate the discussion with the concept of \gls{se} and then expand into \gls{ee}.

\subsubsection{Spectral Efficiency}

To assess the wireless system performance, it is crucial to evaluate the capacity for transmitting bits per second per Hertz, namely as \gls{se}, a fundamental metric that measures the efficiency of information transmission through a channel. The \gls{se} reflects the average number of bits transmitted per channel use across varying fading conditions, serving as a common metric for evaluating communication system capabilities. Since for the \gls{ris}-assisted systems, the channel is a generalization of the conventional \gls{m-mimo} systems, the \gls{se} can be equally represented as the following
\begin{equation}
{\rm SE}_k = \log_2(1+{\rm SINR}_k), \qquad \left[\frac{\rm bit}{\rm s\cdot Hz}\right]
\end{equation}
where the \gls{sinr} of the $k$-th \gls{ue}, ${\rm SINR}_k$, can be expressed as
\begin{equation} \label{eq:SINR}
    {\rm SINR}_k = \frac{|\boldsymbol{w}_k^H \mathbf{h}_k|^2}{ \sum_{j=1, j\neq K}^{K} |\boldsymbol{w}_j \mathbf{h}_k|^2 + |\boldsymbol{z}^H \boldsymbol{\Phi} \boldsymbol{h}_{2,k}|^2 + |n_k|^2 }.
\end{equation}
Finally, we can define the system's normalized \gls{sr} as the following
\begin{equation}
    R = \sum_{k=1}^K \log_2\left(1 + \frac{|\boldsymbol{w}_k^H \mathbf{h}_k|^2}{ \sum_{j=1, j\neq K}^{K} |\boldsymbol{w}_j \mathbf{h}_k|^2 + |\boldsymbol{z}^H \boldsymbol{\Phi} \boldsymbol{h}_{2,k}|^2 + |n_k|^2 }\right).
\end{equation}

\subsubsection{Energy Efficiency}
The metric for evaluating a system's sustainable capacity is referred to as \gls{ee}. This metric essentially quantifies the number of bits that can be reliably transmitted per unit of energy, expressed as:
\begin{equation}\label{eq:EE}
\mathrm{EE} = \frac{\mathrm{ Normalized ~ Sum ~  Rate ~ [bit/s/Hz]}}{\mathrm{{System} ~  Power~Consumption~[W]}} = \frac{\sum_k {\rm SE}_k}{P_{{\rm total}}} \qquad {\left[\frac{\rm bit}{\rm Joule \cdot Hz}\right]}
\end{equation}
where $P_{\rm total}$ is the total power consumed by the communication system as detailed in Remark \ref{rmk:2} and formulated in Subsection \ref{sec:powerconsump} in the next.

{\bf Remark 1} {The \gls{ee} optimization as a function of the amount of transmit \gls{rf} power in \eqref{eq:EE} involves a ratio of a convex function in the numerator and an affine function in the denominator. Hence, one can formulate the \gls{ee} optimization problem as a single-ratio problem, briefly revisited in subsection \ref{sec:FracProg}.}

{\bf Remark 2} {The power consumption for \gls{ris}-assisted \gls{m-mimo} systems requires more detailed consideration than traditional \gls{m-mimo} systems since although the \gls{ris} technology achieves low power consumption, its consumption is not entirely negligible and should be thoroughly examined. Therefore, in the next subsection \ref{sec:powerconsump} we further detail the adopted system power consumption model.}\label{rmk:2}

\subsubsection{System's Power Consumption Model}\label{sec:powerconsump}

Different from most prior works, which mostly consider entirely passive \gls{ris}, here, we adopt a more practical, realistic energy consumption model by taking into account two primary components: the power consumption of the \gls{bs} and the power consumption related to the active \gls{ris}. Generically, the total power consumption model can be written as
\begin{equation}
    P_{{\rm total}} = P_{{\rm BS}} + P_{{\rm RIS}}.
\end{equation}

The power consumption at the \gls{bs} is related directly to three main factors: the constant power required for \gls{bs} operation, $i.e.$, site-cooling, control signaling, load-independent power of backhaul infrastructure, baseband processor, etc; the power necessary to run the circuit components attached to each antenna, such as converters, mixers, filters, etc; and the \gls{rf} transmit power allocated to the \glspl{ue} \cite{7031971}, resulting in 
\begin{equation} \label{eq:pbs}
P_{{\rm BS}} =   P_{0,{\rm BS}} + M P_{{\rm M}} + \sum_{k=1}^K \varrho ||\boldsymbol{w}_k||^2,
\end{equation}
where $\varrho$ denotes the inefficiency of the transmit power amplifier.

On the other hand, the \gls{ris} needs a controller\footnote{The controller includes a \gls{fpga} and drive circuits.}, which is required for receiving the external signals, processing data, and programming, as well as configuring the \gls{ris} unit cells. Overall, the total power consumption dissipated to operate the \gls{ris} consists of three
parts; {\bf a}) one is the static power consumption generated by the \gls{fpga} control board and drive circuits, namely $P_{\rm CB}$ \cite{wang2024reconfigurable,Pei_2021}; {\bf b}) the power consumption by the \gls{ris} unit cells, being, the power necessary for controlling the impedance of each element via amplifier and phase turner, in order to configure the phase/amplitude of reflection; and {\bf c}) the active power for \gls{rf} signal amplification. Accordingly, the power consumed at the active \gls{ris} can be expressed as
\begin{equation} \label{eq:pris1}
    P_{{\rm RIS}} = P_{{\rm CB}} + N P_{{\rm N}} + P_{{\rm RIS}}^{{\rm out}} - P_{{\rm RIS}}^{{\rm in}}.  
\end{equation}
where $P_{{\rm RIS}}^{{\rm out}}$ and $P_{{\rm RIS}}^{{\rm in}}$ is the signal \gls{rf} power that arrives at the \gls{ris} and departs from the \gls{ris}, respectively, given by \cite{fotock2023energy}
\begin{equation}
    P_{{\rm RIS}}^{{\rm in}} = ||\boldsymbol{z}||^2 + \sum_{k=1}^K ||\boldsymbol{w}_k^H \boldsymbol{H}_1||^2 = \textrm{tr}(\boldsymbol{Z} \boldsymbol{Z}^H)  + \sum_{k=1}^K \textrm{tr}(\boldsymbol{U}_k \boldsymbol{U}_k^H),
\end{equation}
\begin{equation}
    P_{{\rm RIS}}^{{\rm out}} = || \boldsymbol{\Phi} \boldsymbol{z} ||^2 + \sum_{k=1}^K || \boldsymbol{w}_k^H \boldsymbol{H}_1\boldsymbol{\Phi}||^2 =  \textrm{tr}\left( \boldsymbol{Z} \boldsymbol{v} \boldsymbol{v}^H \boldsymbol{Z}^H \right)   + \sum_{k=1}^K \textrm{tr}(\boldsymbol{U}_k \boldsymbol{v} \boldsymbol{v}^H \boldsymbol{U}_k^H), 
\end{equation}
where $\boldsymbol{Z} \triangleq \textrm{diag}(\boldsymbol{z})$ and $\boldsymbol{U}_k \triangleq \textrm{diag}(\boldsymbol{H}_1^H \boldsymbol{w}_k)$. Thereby, the total \gls{rf} power amplification at the active \gls{ris} can be expressed as
\begin{IEEEeqnarray}{rCl}
    P_{{\rm RIS}}^{{\rm out}} - P_{{\rm RIS}}^{{\rm in}} &=&   \textrm{tr}\left(\boldsymbol{Z}\left(\boldsymbol{v}\boldsymbol{v}^H - \mathbf{I}_N\right)\boldsymbol{Z}^H\right)  + \sum_{k=1}^K \textrm{tr}\left(\boldsymbol{U}_k \left( \boldsymbol{v} \boldsymbol{v}^H - \mathbf{I}_N \right) \boldsymbol{U}_k^H \right)   
    \\
    &=& \textrm{tr}\left( \left(\boldsymbol{v} \boldsymbol{v}^H - \mathbf{I}_N \right) \boldsymbol{Z}^H \boldsymbol{Z} \right) + \textrm{tr} \left( \left(\boldsymbol{v} \boldsymbol{v}^H - \mathbf{I}_N \right) \sum_{k=1}^K\boldsymbol{U}_k^H \boldsymbol{U}_k \right) \nonumber
    \\
    &=& \textrm{tr} \left(\left( \boldsymbol{v}\boldsymbol{v}^H - \mathbf{I}_K\right) \boldsymbol{Q}\right),
    \nonumber \\ 
    &=& \boldsymbol{v}^H\boldsymbol{Q}\boldsymbol{v} - \textrm{tr}(\boldsymbol{Q}) \nonumber
\end{IEEEeqnarray}
where 
\begin{equation}
    \boldsymbol{Q} \triangleq \boldsymbol{Z}^H \boldsymbol{Z} + \sum_{k=1}^K \boldsymbol{U}_k^H\boldsymbol{U}_k
\end{equation}
Therefore, Eq. \eqref{eq:pris1} also can be written as a function of $\boldsymbol{v}$ as following\footnote{Herein, we assume that the \gls{ris} is unable to absorb the incident power, although it is possible by means of energy harvesting circuits embedded at the \gls{ris}. Therefore, $P_{\textrm{RIS}}^{\rm out} \geq P_{\textrm{RIS}}^{\rm in}$ always hold, with the equality when $|v_n| =1$, $\forall n$ $\in \mathcal{N}$.}  
\begin{equation}
    P_{\rm RIS} = 
    P_{{\rm CB}} +  N P_{{\rm N}} + \boldsymbol{v}^H\boldsymbol{Q} \boldsymbol{v} - \textrm{tr}\left(\boldsymbol{Q} \right).
\end{equation}
A detailed discussion on the static power consumption generated by the
\gls{fpga} board and drive circuits $P_{\rm CB}$ is found in \cite{wang2024reconfigurable,Pei_2021}.

\section{Optimization Techniques} \label{sec:optzTech}

In order to fully clarify the proposed optimization strategy for tackling the \gls{ee} active \gls{ris}-aided \gls{m-mimo} optimization problem, it is crucial to delve into some key optimization techniques that act in our proposed solution.
These techniques play essential roles in our approach to addressing the investigated problem, and an entire comprehension of their concepts is essential for a careful understanding of the proposed solution. In the following, we provide a detailed review of these techniques, including \gls{ldt} and fractional 
single and multiple-ratio problems.

\subsection{Lagrangian Dual Transform}

{In addressing the challenge posed by the sum-of-logarithms problem, specific techniques prove to be valuable in easing the management of the {optimization} problem, thereby enabling the possibility of finding a closed-form, low-complexity solution. Given that communication system-related problems frequently involve logarithmic function summation, finding any form of solution can be exceedingly challenging as the presence of logarithmic functions hinders the analytical tractability and, consequently, optimization. As a result, {an optimization technique introduced by \cite{8310563}, namely \gls{ldt}}  leverages the Lagrangian duality to externalize the argument of logarithm functions to the outside, eliminating the dependence of the variable of interest with the logarithmic functions, resulting in easier management and analytical tractability of the  problem. To further understand the \gls{ldt}, let us define the following target maximization problem:
\begin{IEEEeqnarray}{rCl}\label{eq:waux1}
    &\underset{ \mathbf{x} }{ \mathrm{maximize}}& \qquad \sum_{\ell=1}^{L} \log\left(1 + \frac{f_\ell(\mathbf{x})}{g_\ell(\mathbf{x})}\right),
    \\
    &\mathrm{subject\,to}& \qquad h_i(\mathbf{x}) \leq 0, \quad i \in \{1,\dots,I\} \IEEEyessubnumber \label{eq:waux1111}
\end{IEEEeqnarray}
where $f_\ell(\mathbf{x}):$ $\mathbb{C}^N \rightarrow \mathbb{R}_{+}$ is a non-negative function, and $g_\ell(\mathbf{x}):$ $\mathbb{C}^N \rightarrow \mathbb{R}_{+}^{*}$ is a positive function, $\forall \ell$ $\in \{1,\dots,L\}$. It is important to observe that the ratio $f_\ell(\mathbf{x})/g_\ell(\mathbf{x})$ can have a physical interpretation as \gls{sinr} in our context. The problem \eqref{eq:waux1} is assumed to be a non-convex problem, with a non-convex constraint \eqref{eq:waux1111}. According to \cite{8314727}, the problem \eqref{eq:waux1} can be transformed in the following equivalent problem
\begin{IEEEeqnarray}{rCl}\label{eq:waux2}
    &\underset{ \mathbf{x}, \, \boldsymbol{\alpha} }{ \mathrm{maximize}} & \qquad    \sum_{\ell=1}^L \log(1 + \gamma_\ell) - \sum_{\ell=1}^L \gamma_\ell + \sum_{\ell=1}^L (1+\gamma_\ell)\frac{ f_\ell(\mathbf{x})}{f_\ell(\mathbf{x}) + g_\ell(\mathbf{x})},
    \\
    \nonumber 
    \\
    & \mathrm{subject \, to} & \qquad h_i(\mathbf{x}) \leq 0, \qquad \forall i \in \{1,\dots,I\}  \IEEEyessubnumber 
    \\
    &&\qquad  \gamma_\ell \geq 0, \qquad \forall \ell \in \{1,\dots,L\}. \IEEEyessubnumber
\end{IEEEeqnarray}
where $\boldsymbol{\gamma}^{(t)} = [\gamma_1^{(t)},\gamma_2^{(t)},\dots,\gamma_L^{(t)}]^{T}$ is an auxiliary variable, which is iteratively updated, introduced for each ratio term $f_\ell(\mathbf{x}^{(t-1)})/g_\ell(\mathbf{x}^{(t-1)})$. It's important to note that the two problems, \eqref{eq:waux1} and \eqref{eq:waux2}, are equivalent. In other words, at the convergence, the solution $\mathbf{x}$ to \eqref{eq:waux1} is identical to the solution to \eqref{eq:waux2}, and their respective optimal objective values are also equal \cite[III-B]{8310563}. 
Furthermore, it is worth highlighting that, according to the \gls{kkt} conditions, we can obtain the following
\begin{IEEEeqnarray}{rCl} \label{eq:aux22}
    &\frac{\partial \left( \displaystyle \sum_{\ell=1}^L \log(1 + \gamma_\ell^{(t+1)}) - \displaystyle \sum_{\ell=1}^L \gamma_\ell^{(t+1)} + \displaystyle \sum_{\ell=1}^L \dfrac{(1+\gamma_\ell^{(t+1)}) f_\ell(\mathbf{x}^{(t)})}{f_\ell(\mathbf{x}^{(t)}) + g_\ell(\mathbf{x}^{(t)})} \right) }{\partial \gamma_\ell }  = 0&
    \\
    &\gamma_\ell^{\star (t+1)} = \frac{f_\ell(\mathbf{x}^{(t)})}{g_\ell(\mathbf{x}^{(t)})}& \label{eq:aux222}
\end{IEEEeqnarray}
This equivalence allows for a seamless transition between the two problem formulations while preserving the optimality.
}

\subsection{Fractional Programming}\label{sec:FracProg}

Fractional programming theory is the branch of optimization theory concerned with the properties and optimization of fractional
functions, \emph{i.e.}, a ratio of two functions that are generally nonlinear. They can be found in several areas and are generically subdivided into two classes: single-ratio problems and multiple-ratio problems. Below, we review techniques designed to address each of these classes and their respective solution methodologies.

\subsubsection*{Single-Ratio Problems}
There are many transforms that deal consistently with this type of problem; the most classical is named \textit{Dinkelbach's} Transform \cite{dinkel}, \cite{8314727}, \cite{8310563}. For a better understanding, let us define the following optimization problem. 
\begin{IEEEeqnarray}{rCl}  \label{eq:waux3}
    &\underset{ \mathbf{x} }{ \mathrm{maximize}}&  \quad \frac{f(\mathbf{x})}{g(\mathbf{x})},
    \vspace{0.2cm}
    \\
    & \mathrm{subject\,to}& \quad h_i(\mathbf{x}) \leq 0, \qquad \forall i \in \{1,\dots,I\} \IEEEyessubnumber
\end{IEEEeqnarray}
where $f(\mathbf{x}): \mathbb{C}^{N} \rightarrow \mathbb{R}_{+}$ and $g(\mathbf{x}): \mathbb{C}^{N} \rightarrow \mathbb{R}^{*}_{+}$ are non-negative function and positive function, respectively. 
The conventional approach for dealing with this \gls{fp} is decoupling the numerator and denominator and treating it jointly. The \textit{Dinkelbach's} transform reformulates the single-ratio problem \eqref{eq:waux3} as the following
\begin{IEEEeqnarray}{rCl} \label{eq:waux4}
    &\underset{ \mathbf{x}, \eta }{ \mathrm{maximize}}& \qquad f(\mathbf{x}) - \eta g(\mathbf{x}),
    \\
    & \mathrm{subject\,to}& \qquad h_i(\mathbf{x}) \leq 0, \qquad \forall i \in \{1,\dots,I\} \IEEEyessubnumber 
\end{IEEEeqnarray}
where $\eta$ is an auxiliary variable that is iteratively updated by
\begin{equation} \label{eq:waux5}
    \eta^{(t)} = \frac{f(\mathbf{x}^{(t-1)})}{g(\mathbf{x}^{(t-1)})},
\end{equation}
where $t$ is the iteration index.

\subsubsection*{Multiple-Ratio Problems}
Although the classic \textit{Dinkelbach's} transform \cite{dinkel} works well for single-ratio problems, they cannot be easily extended to the multiple-ratio problems, since for single-ratio problems, the optimal solution is the same for the original \gls{fp} and the transformed problem, but not the value of the objective function of both. To solve the multi-ratio problem, several different techniques have been proposed, such as \textit{Quadratic Transform} \cite{8310563}, and \cite{Jong2012AnEG}. Herein, we focus on the technique developed by \cite{Jong2012AnEG}. Let us define the following multi-ratio problem:
\begin{IEEEeqnarray}{rCl} \label{eq:waux6}
     &\underset{ \mathbf{x} }{ \mathrm{maximize}}&  \qquad \sum_{\ell=1}^{L} \frac{f_\ell(\mathbf{x})}{g_\ell(\mathbf{x})},
     \\
     &\mathrm{subject\,to}& \qquad h_i(\mathbf{x}) \leq 0, \qquad i \in \{1,\dots,I\} \IEEEyessubnumber
\end{IEEEeqnarray}
where $f_\ell(\mathbf{x}): \mathbb{C}^{N} \rightarrow \mathbb{R}_+$ and $g_\ell(\mathbf{x}): \mathbb{C}^{N} \rightarrow \mathbb{R}_{+}^{*}$, $\forall \ell$ $\in \{1,\dots,{L}\}$.
In \cite{Jong2012AnEG}, the authors proposed an iterative solution in order to solve problem \eqref{eq:waux6}, where the equivalent problem can be given as
\begin{IEEEeqnarray}{rCl} \label{eq:waux7}
    &\underset{ \mathbf{x}, \mathbf{u}, \boldsymbol{\beta} }{ \mathrm{maximize}}&  \qquad \sum_{\ell=1}^{L} u_\ell\left(f_\ell(\mathbf{x}) - \beta_\ell g_\ell(\mathbf{x}) \right),
    \\ 
    &\mathrm{subject\,to}& \qquad h{_i}(\mathbf{x}) \leq 0, \qquad i \in \{1,\dots,I\} \IEEEyessubnumber
\end{IEEEeqnarray}
where $\mathbf{u}$ and $\boldsymbol{\beta}$ are auxiliary variables which are iteratively updated as 
\begin{equation} \label{eq:waux8}
    u_\ell^{(t+1)} = \frac{1}{g{_\ell}(\mathbf{x}^{(t)})}, \quad \forall \ell=1,\dots,L,
\end{equation}
\begin{equation} \label{eq:waux9}
    \beta_\ell^{(t+1)} = \frac{f{_\ell}(\mathbf{x}^{(t)})}{g_{\ell}(\mathbf{x}^{(t)})}, \quad \forall \ell=1,\dots,L.
\end{equation}

\section{Problem Formulation} \label{sec:ProblemFormulation}

In view of addressing all the aforementioned arguments, with regard to \gls{ee} exposed in Section \ref{sec:intro}, we formulate an optimization problem whose main objective is optimizing the reflecting coefficients of an active \gls{ris} in order to assess the necessary number of \gls{ris} elements that should be selected to outperform the entirely passive \gls{ris} in terms of \gls{ee}. The idea is to reduce the number of operating elements in the \gls{ris} in view of reducing their power consumption while increasing the total \gls{sr} by amplifying the incoming signal intended for the \glspl{ue}. Hence, such an objective can be formulated as the following optimization problem, denoted as $\mathcal{P}_0$:
\begin{IEEEeqnarray}{rCl} \label{eq:pEE}
    \mathcal{P}_0: \quad & \underset{\boldsymbol{v}}{\mathrm{maximize}}&  \qquad \frac{ \displaystyle \sum_{k=1}^K \log_2\left(1 + \dfrac{|\boldsymbol{w}_k^H \boldsymbol{h}_k|^2}{ \sum_{j=1, j\neq K}^{K} |\boldsymbol{w}_j \boldsymbol{h}_k|^2 + |\boldsymbol{z}^H \boldsymbol{\Phi} \boldsymbol{h}_{2,k}|^2 + |n_k|^2 }\right) }{ P_{0,{\rm BS}} + P_{\rm CB} + M P_{{\rm M}} + \sum_{k=1}^K \varrho ||\boldsymbol{w}_k||^2 +  N P_{{\rm N}} + \boldsymbol{v}^H\boldsymbol{Q} \boldsymbol{v} - \textrm{tr}\left(\boldsymbol{Q} \right)  },
    \vspace{0.2cm}
    \\
    & \mathrm{subject\,to}&  \hspace{0.8cm} \textrm{tr}(\boldsymbol{Q}) \leq \boldsymbol{v}^H \boldsymbol{Q} \boldsymbol{v} \leq \mathrm{tr}(\boldsymbol{Q}) + P_{\max}^{{\rm RIS}}, 
    \IEEEyessubnumber \label{eq:c1}
    \\
    && \hspace{0.8cm} |v_n| \leq \alpha_{\max}, \qquad \forall n \in \mathcal{N}. \IEEEyessubnumber \label{eq:c2}
\end{IEEEeqnarray}
Constraint \eqref{eq:c1} considers the maximum amplification of power provided by the active \gls{ris}, while constraint \eqref{eq:c2} considers the maximum amplitude imposed by each element of the \gls{ris}. One should notice that problem $\mathcal{P}_0$ is more challenging than the \gls{ee} optimization with entirely passive \gls{ris} since the \gls{ris} configuration vector also appears in the denominator of the objective function \eqref{eq:pEE}.

\section{Proposed Solution}\label{sec:Solution}
In Subsection \ref{subsec:Alg}, we introduce the proposed algorithm namely \gls{fp}-based Amplitude/Phase Beamforming Design, which utilizes the \gls{fp} optimization technique. Specifically, to enhance the convexity of the optimization problem, this algorithm incorporates auxiliary variables derived from Dilkelbach's Transform, \gls{ldt}, and from the methodology outlined in \cite{Jong2012AnEG}. This integration strategically molds the objective function into a convex form, facilitating effective optimization.

Subsequently, in Subsection \ref{subsec:Complexity}, we conduct a comprehensive analysis of the algorithm's complexity. This examination provides valuable insights, emphasizing the algorithm's efficiency and suitability for real-world applications.

Finally, in Section \ref{sec:Result}, we present a comparative evaluation of the proposed algorithm's performance within typical \gls{ris}-aided \gls{m-mimo} channel and system scenarios. This comparative analysis serves to validate and illustrate the algorithm's efficacy across a range of practical and diverse operating conditions.

\subsection{\gls{fp}-Based Solution Method} \label{subsec:Alg}

Before delving directly into the problem at hand, it is useful to establish a more concise nomenclature. Therefore, firstly, for the sake of tractability, we shall rewrite Equation \eqref{eq:SINR} as follows
\begin{equation}
    \textrm{SINR}_k =  \frac{|\boldsymbol{w}_k^H \boldsymbol{h}_k|^2}{ \sum_{j=1, j\neq K}^{K} |\boldsymbol{w}_j \boldsymbol{h}_k|^2 + | \boldsymbol{z}^H \boldsymbol{\Phi} \boldsymbol{h}_{2,k} |^2 + |n_k|^2}  =  \frac{\boldsymbol{v}^H\boldsymbol{A}_k \boldsymbol{v}}{\boldsymbol{v}^H\boldsymbol{B}_k \boldsymbol{v} + |n_k|^2} ,
\end{equation}
 where with few analytical manipulations, we obtain that 
 \begin{IEEEeqnarray}{rCl}
     \boldsymbol{A}_k &\triangleq& \boldsymbol{H}_{2,k}^H \boldsymbol{H}_1^H \boldsymbol{w}_k \boldsymbol{w}_k^H \boldsymbol{H}_1 \boldsymbol{H}_{2,k}\,,
     \\
     \boldsymbol{B}_k &\triangleq&  \boldsymbol{H}_{2,k}^H \boldsymbol{z} \boldsymbol{z}^H \boldsymbol{H}_{2,k}  +  \sum_{j=1,j\neq k}^{K}\boldsymbol{H}_{2,k}^H\boldsymbol{H}_1^H \boldsymbol{w}_j \boldsymbol{w}_j \boldsymbol{H}_1 \boldsymbol{H}_{2,k}\,,
 \end{IEEEeqnarray}
with $\mathbf{H}_{2,k} = \textrm{diag}(\mathbf{h}_{2,k})$. This manipulation enables us to effortlessly address our problem, thereby facilitating the derivation of subsequent equations.
Rewriting problem \eqref{eq:pEE}, we have the following 
\begin{IEEEeqnarray}{rCl} \label{eq:Prew}
     \mathcal{P}_0: \quad & \underset{\boldsymbol{v}}{\mathrm{maximize}}&  \qquad \frac{ \displaystyle \sum_{k=1}^K 
    \log_2\left(1 + \frac{\boldsymbol{v}^H\boldsymbol{A}_k \boldsymbol{v}}{ \boldsymbol{v}^H\boldsymbol{B}_k \boldsymbol{v} +  |n_k|^2 } \right) }{ \hat{P} +  \boldsymbol{v}^H\boldsymbol{Q}\boldsymbol{v}},
    \vspace{0.2cm}
    \\
    & \mathrm{subject\,to}&  \qquad \eqref{eq:c1}, ~ \eqref{eq:c2} \nonumber,
\end{IEEEeqnarray}
where we define $\hat{P} \triangleq \sum_{k=1}^K \varrho ||\boldsymbol{w}_k||^2 + P_{0,\rm{BS}} + P_{\rm CB}  + M P_{\rm{M}} + N P_{{\rm N}} - \rm{tr}(\mathbf{Q})$. One can see that Problem \eqref{eq:Prew} is a non-convex problem, with a non-convex constraint \eqref{eq:c1}. Furthermore, we also should notice that \eqref{eq:Prew} is a single ratio problem.  Therefore, prior to solving the optimization process, a pre-processing step should be applied to the objective function due to its intractable fractional form. To decouple the numerator and denominator of \eqref{eq:Prew}, we employ the classic Dinkelbach's methodology \cite{dinkel}. Specifically, by introducing an auxiliary variable $\eta$,  according to subsection \ref{sec:FracProg}, the original problem can be equivalently reformulated as
\begin{IEEEeqnarray}{rCl} \label{eq:p1}
    \mathcal{P}_1: \quad & \underset{\boldsymbol{v}, \eta }{\mathrm{maximize}}&  \qquad  \displaystyle \sum_{k=1}^K 
    \log_2\left(1 + \frac{\boldsymbol{v}^H\boldsymbol{A}_k \boldsymbol{v}}{ \boldsymbol{v}^H\boldsymbol{B}_k \boldsymbol{v} +  |n_k|^2} \right) - \eta \left(\hat{P} +  \boldsymbol{v}^H\boldsymbol{Q}\boldsymbol{v}\right),
    \vspace{0.2cm}
    \\
    & \mathrm{subject\,to}&  \qquad \eqref{eq:c1}, ~ \eqref{eq:c2}\nonumber ,
\end{IEEEeqnarray}
where the optimal $\eta^*$ in the $\ell$-th iteration can be directly computed as
\begin{equation} \label{eq:eta}
    \eta^{\star{(\ell)}} =     \dfrac{\displaystyle \sum_{k=1}^K \log_2\left(1 + \dfrac{\boldsymbol{v}^{{(\ell-1)} H}\boldsymbol{A}_k \boldsymbol{v}^{{(\ell-1)}}}{ \boldsymbol{v}^{{(\ell-1)}H}\boldsymbol{B}_k \boldsymbol{v}^{{(\ell-1)}} +  \sigma^2}\right) }{{\hat{P} +  \boldsymbol{v}^{{(\ell-1)} H}\boldsymbol{Q}\boldsymbol{v}^{{(\ell-1)}}}}  .
\end{equation}
Thus, we should seek to solve $\mathcal{P}_1$, for fixed $\eta$ (at the $\ell$-th iteration).  In this way, we should see that the objective function of $\mathcal{P}_1$ is still non-convex. Moreover, it is composed of a scaled version of a multiple-ratio function; therefore, applying the \gls{ldt} technique, we can introduce the auxiliary variable $\boldsymbol{\gamma}$, and rewrite $\mathcal{P}_1$ in an equivalent way, obtaining the following optimization problem, which we denote as $\mathcal{P}_2$
\begin{IEEEeqnarray}{rCl}
    \mathcal{P}_2: \quad & \underset{\boldsymbol{v}, \boldsymbol{\gamma}}{\mathrm{maximize}}&  \quad   \displaystyle \sum_{k=1}^K 
    \log_2\left(1 + \gamma_k \right) - \gamma_k +  
  \vphantom{\sum} \frac{(1+\gamma_k)\boldsymbol{v}^H\boldsymbol{A}_k \boldsymbol{v}}{\left(  \boldsymbol{v}^H\left(\boldsymbol{A}_k + \boldsymbol{B}_k \right) \boldsymbol{v} +  |n_k|^2 \right)}   -\eta \left(\hat{P} +  \boldsymbol{v}^H\boldsymbol{Q}\boldsymbol{v}\right), \nonumber
 \\
 & \mathrm{subject\,to}&  \qquad \eqref{eq:c1}, \eqref{eq:c2}. \nonumber
\end{IEEEeqnarray}
The $\gamma$ that maximizes $\mathcal{P}_2$ at the $\ell$-th iteration is given by 
\begin{equation}
\gamma_k^{\star {(\ell)} } =\frac{\boldsymbol{v}^{{(\ell-1)} H} \boldsymbol{A}_k \boldsymbol{v}^{{(\ell-1)}}}{\boldsymbol{v}^{{(\ell-1)} H} \boldsymbol{B}_k \boldsymbol{v}^{{(\ell-1)}} + \sigma^2}, \qquad \forall k \in \mathcal{K}. \label{eq:gamma}
\end{equation}

By proceeding, following methodology of \cite{Jong2012AnEG}, and introducing the following auxiliary variables, $\boldsymbol{\beta}$, and $\boldsymbol{u}$, we can rewrite $\mathcal{P}_2$ as
\begin{IEEEeqnarray}{rCl}
    \mathcal{P}_2^{'}: \quad & \underset{\boldsymbol{v}, \boldsymbol{u}, \boldsymbol{\beta}}{\mathrm{maximize}}&  \quad  \displaystyle \sum_{k=1}^K 
    \log_2\left(1 + \gamma_k \right) - \gamma_k +  
 u_k \left [ \vphantom{\sum} {(1+\gamma_k)\boldsymbol{v}^H\boldsymbol{A}_k \boldsymbol{v}} -\beta_k\left(  \boldsymbol{v}^H\left(\boldsymbol{A}_k + \boldsymbol{B}_k \right) \boldsymbol{v} +  |n_k|^2 \right) \right] \nonumber
 \\
 && \hspace{5cm} -\eta \left(\hat{P} +  \boldsymbol{v}^H\boldsymbol{Q}\boldsymbol{v}\right),
 \\
 & \mathrm{subject\,to}&  \qquad \eqref{eq:c1}, \eqref{eq:c2}. \nonumber
\end{IEEEeqnarray}
The $\beta_k$ and $u_k$ that maximize $\mathcal{P}_2^{'}$ at the $\ell$-th iteration can be obtained by the \gls{kkt} conditions which are given as: 
\begin{IEEEeqnarray}{rCl}
    \beta_{k}^{(\ell) \star} &=&
    \frac{(1+\gamma_k^{(\ell)})\boldsymbol{v}^{ {(\ell-1)} H}\boldsymbol{A}_k \boldsymbol{v}^{{(\ell-1)}}}{\boldsymbol{v}^{{(\ell-1)} H} \left(\boldsymbol{A}_k +  \boldsymbol{B}_k\right) \boldsymbol{v}^{{(\ell-1)}} + |n_k|^2}, \qquad \forall k \in \mathcal{K}, \label{eq:beta} 
    \\
    \quad u_k^{(\ell) \star} &=& \frac{1}{\boldsymbol{v}^{{(\ell-1)}H}\left(\boldsymbol{A}_k +  \boldsymbol{B}_k\right) \boldsymbol{v}^{{(\ell-1)}} + |n_k|^2}, \qquad \forall k \in \mathcal{K}.  \label{eq:u}
\end{IEEEeqnarray}
Rewriting $\mathcal{P}_2^{'}$ in a more compact form, we arrive at the following equivalent problem
\begin{IEEEeqnarray}{rCl}
\mathcal{P}_3: \quad & \underset{\boldsymbol{v}}{\mathrm{maximize}}&  \qquad   \boldsymbol{v}^H \boldsymbol{C} \boldsymbol{v},
    \vspace{0.2cm}
    \\
    & \mathrm{subject\,to}&  \qquad \eqref{eq:c1}, ~ \eqref{eq:c2}, \nonumber 
\end{IEEEeqnarray}
where 
\begin{equation}
\boldsymbol{C}^{(\ell)} = - \eta^{(\ell)} \boldsymbol{Q} + \sum_{k=1}^K u_k^{(\ell)}(1+\gamma_k^{(\ell)}) \boldsymbol{A}_k - u_k^{(\ell)} \beta_k^{(\ell)} \left(\boldsymbol{A}_k + \boldsymbol{B}_k  \right).
\end{equation}

\vspace{2mm}
The problem $\mathcal{P}_3$ can be rewritten in an alternative manner as follows:
\begin{IEEEeqnarray}{rCl} \label{eq:Pfinal}
\mathcal{P}_4: \quad & \underset{\boldsymbol{V}}{\mathrm{maximize}}&  \qquad   {\rm tr}\left( \boldsymbol{C} \boldsymbol{V} \right),
    \vspace{0.2cm}
    \\
    & \mathrm{subject\,to}&  \qquad \textrm{tr} \left( \boldsymbol{C} \boldsymbol{V} \right) \leq \mathrm{tr}(\boldsymbol{Q}) + P_{\max}^{{\rm RIS}}, \IEEEyessubnumber
    \\
    && \qquad \textrm{tr}  \left( \boldsymbol{C} \boldsymbol{V} \right) \geq \textrm{tr} \left( \boldsymbol{Q} \right) \IEEEyessubnumber
    \\
    && \qquad [\boldsymbol{V}]_{n,n} \leq \alpha_{\max}^2, \quad \forall n \in \mathcal{N} \IEEEyessubnumber
    \\
    && \qquad {\rm rank}(\boldsymbol{V}) = 1, \IEEEyessubnumber \label{eq:c3}
\end{IEEEeqnarray}
where the $\boldsymbol{v}^H\boldsymbol{C}\boldsymbol{v} = \textrm{tr}(\boldsymbol{C} \boldsymbol{V})$ property is utilized, with $\boldsymbol{V} = \boldsymbol{v}\boldsymbol{v}^H$. In particular, both the objective function and constraints in $\mathcal{P}_4$, given by Eq. \eqref{eq:Pfinal}, are linear in the matrix $\boldsymbol{V}$ \cite{5447068}. %

Notice that the constraint \eqref{eq:c3} is a non-convex constraint; however, we can relax it in order to obtain a sub-optimal solution for problem $\mathcal{P}_4$.  In summary, the optimization problem $\mathcal{P}_4$ can be globally solved by standard fractional programming tools, such as CVX \cite{cvx}.  If the obtained solution of $\boldsymbol{V}$  has a unit rank, then $\boldsymbol{V}$ is also feasible for the original problem. Otherwise, a feasible solution can be
obtained by randomization techniques \cite{5447068, 9676676}. 

Based on the above derivations, the overall \gls{fp}-based Amplitude/Phase Beamforming design algorithm is summarized in Algorithm \ref{alg:FPbeam}. By appropriately initializing the variable $\boldsymbol{v}$ and the \gls{bs} transmit power $\boldsymbol{p}$, the \gls{ris} reflection/amplification matrix $\boldsymbol{\Phi} = {\rm diag}(\boldsymbol{v})$ is iteratively updated, until the
\gls{ee} of the \gls{ris}-assisted \gls{m-mimo} system converges.

\begin{algorithm}[htbp!] 
\caption{\bf \gls{fp}-based Amplitude/Phase Beamforming Design}\label{alg:FPbeam}
\begin{algorithmic}
\State \textbf{Input:} ${\boldsymbol{p}}$, $\mathbf{A}_k$, $\mathbf{B}_k$, $\forall k$ $\in \{1,\dots,K\}$
\State Set $\ell = 1$;
\State Set feasible value for ${\boldsymbol{v}^{(\ell-1)}}$;
\State Set $\eta^{(\ell-1)}=0$;
\Repeat
\State \textbf{Step 1:} Compute $\eta^{\star (\ell)}$ by \eqref{eq:eta};
\State \textbf{Step 2:} Update $\boldsymbol{\gamma}^{\star (\ell)}$ by \eqref{eq:gamma};
\State \textbf{Step 3:}  Compute $\boldsymbol{\beta}^{\star (\ell)}$ by \eqref{eq:beta}; 
\State \textbf{Step 4:}  Update $\boldsymbol{u}^{\star (\ell)}$ by \eqref{eq:u}; 
\State \textbf{Step 5:} Obtain $\boldsymbol{v}^{(\ell)}$ by solving $\mathcal{P}_4$;
\State \textbf{Step 6:} $\ell = \ell + 1$;
\vspace{0.2cm}
\Until $|\eta^{\star (\ell)} - \eta^{\star (\ell-1)}| < \epsilon$
\State \textbf{Output:} $\boldsymbol{v}^\star = [\alpha_1^{ \star} e^{-j\theta_1^{ \star}},\alpha_2^{\star} e^{-j\theta_2^{\star}},\dots,\alpha_N^{\star} e^{-j\theta_N^{ ^\star}}]^H$
\end{algorithmic}
\label{alg:phase-optz}
\end{algorithm}

\subsection{Complexity} \label{subsec:Complexity}
So far, we have completed the design of active \gls{ris} phase shift matrix $\boldsymbol{\Phi}$. For clarity of this scheme procedure, we summarize the main ideas of the entire proposed \gls{fp}-based amplitude/phase beamforming design algorithm as follows. First, by fixing $\eta$, we can obtain $\boldsymbol{\gamma}$, $\boldsymbol{\beta}$, and $\boldsymbol{u}$, with closed-form expressions, and the RIS phase-shift matrix $\boldsymbol{v}$ by solving $\mathcal{P}_4$, as illustrated on Algorithm \ref{alg:FPbeam}. The alternating iterating procedure is repeated among four variables until
the termination condition is reached.
Herein, we describe the complexity of the proposed iterative algorithm. It is known that, the complexity to update $\boldsymbol{\gamma}$, $\boldsymbol{\beta}$, $\boldsymbol{u}$, and solving $\mathcal{P}_4$ are $\mathcal{O}(2KN^2)$, $\mathcal{O}(2K N^2)$, $\mathcal{O}(KN^2)$, and $\mathcal{O}(N^2)$ \cite{5447068} respectively. Besides, the $\eta$ updating requires $\mathcal{O}(3KN^2)$. Therefore, the overall complexity is given by $\mathcal{O}\left( 8LKN^2 \ln(\frac{1}{\epsilon}) \right)$
\gls{flops}, where $L$ denotes the number of loop iterations. {In addition, it is worthy to notice that the complexity of the Gaussian randomization method is of $\mathcal{O}(N^2)$ per draw \cite{1673411}.}

\section{Numerical Results} \label{sec:Result}

Herein, in this section we illustrate simulation results in view to demonstrate the effectiveness of the proposed \gls{fp}-based amplitude/phase beamforming design algorithm, where the active \gls{ris} amplitude/phase vector is obtained by running Algorithm \ref{alg:phase-optz}, where we solve $\mathcal{P}_4$ via convex optimization with CVX solver, in MATLAB. For the simulation setup, we assume that the \glspl{ue} are positioned in a circle with a radius of $10~m$ from the center $(30~m,30~m)$,  the \gls{bs} is located at $(-10~m, 5~m)$ and the \gls{ris} is located at the origin $(0~m,0~m)$. For the \gls{bs}\gls{ris} link, we defined $\beta_0^{{\rm BR}} = 1e^{-3}$ and $\lambda^{\rm BR} = 2$, while for the \gls{ris}-\glspl{ue} link, we assume $\beta_0^{{\rm RU}} = 1e^{-3}$ and $\lambda^{\rm RU} = 2.5$. 
Unless specified otherwise, the remainder of the parameters are listed in Table \ref{tab:m-mimo-ris-simulation-parameters}. Aiming for a fair comparison, we assume the same power budget for entirely passive \gls{ris} and active \gls{ris}, where the maximum power for \gls{ris} amplification, $P_{\max}^{{\rm RIS}}$, is given by $P_{\max}^{{\rm RIS}} = \tau_{{\rm RIS}} P_{{\rm TX}}$, where $\tau_{\rm RIS}$ $\in [0,1]$. 

\begin{table}[!h]
    \caption{List of simulation parameters.
    \label{tab:m-mimo-ris-simulation-parameters}}{
    \normalsize
    \centering
    \begin{tabular}{@{}|rl|@{}}
        \hline
        \textbf{Parameter} & \textbf{Description}\\
          \hline
        $P_{\textsc{tx}}  =  [10:5:50]$ dBm &
        Maximum power budget at \gls{bs} \\
        $\sigma_k^2 = -95$ dBm & Noise variance at \glspl{ue}, $\forall k$ $\in \mathcal{K}$   \\
        $\sigma_{\rm \gls{ris}}^2 = -80$ dBm & Noise variance at \gls{ris}   \\
        $K = 5$ & Numbers of \gls{ue}s \\
         $M = 32$ & Number of antennas at \gls{bs} \\
        $N =  64$ & Number of reflecting meta-surfaces elements \\
        $\varrho$ = 1.2 & Power amplifier inefficiency  \\
        $P_{0,{\rm BS}}$ = 9 dBW & Fixed power consumption at \gls{bs}   \\
        $P_{\rm{M}}$ = 1 W & Power consumed by a \gls{bs} transceiver chain\\
        $P_{\rm{CB}}$ = 4.8 W & Power consumed by control board\\
        $P_{\rm{N}}$ = 10 dBm & Power of \gls{ris} element configuration   \\
        $\alpha_{\max} = 10$ & Maximum amplitude gain \\
        $\kappa_1=0.1$ & Rician coefficient for \gls{bs} \gls{ris} link \\
        $\kappa_{2,k}=0.01$ & Rician coefficient for \gls{ris} \glspl{ue} links, $\forall k$ $\in \mathcal{K}$ \\
        $\tau_{\rm RIS} = 0.15$& Coefficient for \gls{ris} amplification \\
        $\mathcal{T} = {500}$ & Realizations \gls{mcs} \\
           \hline
    \end{tabular}
    }
\end{table}

\subsection{Energy Efficiency {\it vs.} Transmit Power Budget}

\begin{figure}[!htbp]
\centering
\includegraphics[width=.69\columnwidth]{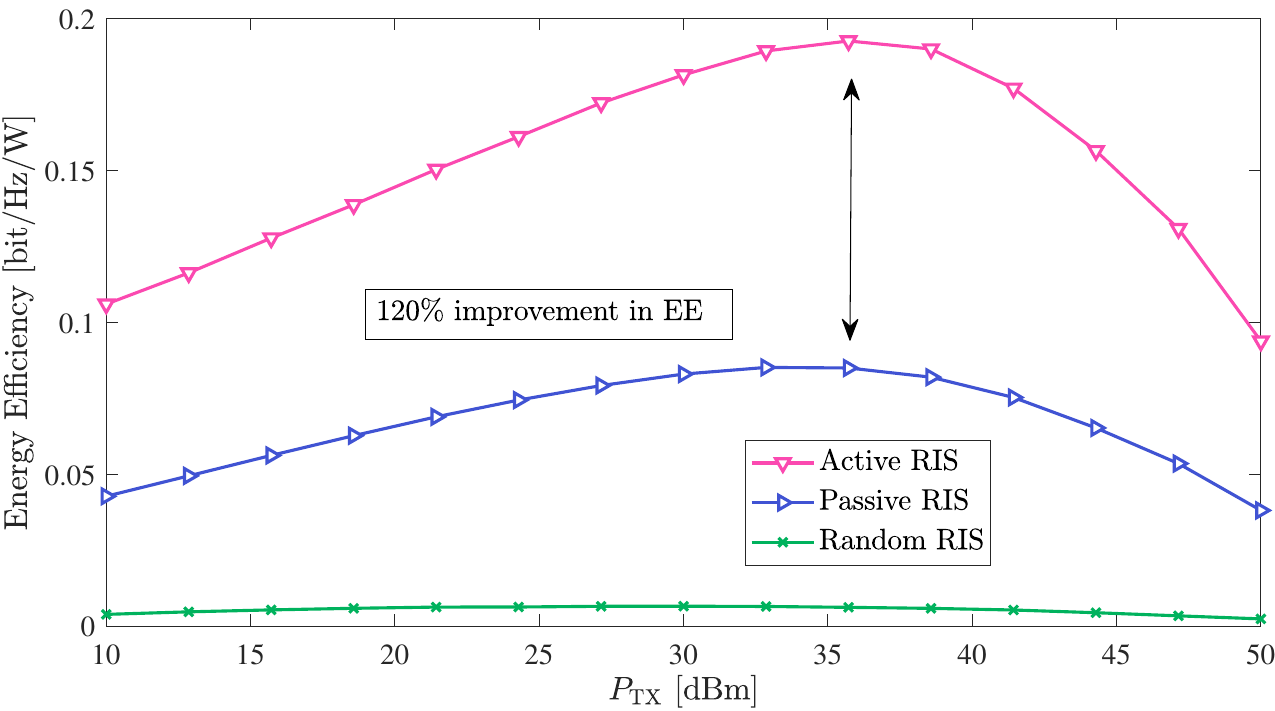}
\caption{Average \gls{ee} {\it vs} the transmit  
power budget [dBm] at the \gls{bs} ($P_{\rm{TX}}$). Performance evaluation of the proposed algorithm for the active \gls{ris} {\it vs} entirely passive \gls{ris} and random phase shift/precoding, with $N=25$. }
\label{fig:avEE}
\end{figure}

Initially, in Fig. \ref{fig:avEE}, we assess and compare the \gls{ee} of both \gls{ris} architecture as a function of transmit power $P_{\rm TX}$. For each point in the curve, the proposed algorithm is employed to optimize the power of reflection of the active \gls{ris} array, while for the passive \gls{ris} we just optimize the phase shift angles. We denote as ``Passive \gls{ris}'' the entirely passive \gls{ris}, ``Active \gls{ris}'' the active \gls{ris},  ``Random \gls{ris}'' as the entirely passive \gls{ris} with random precoding and random \gls{ris} reflecting coefficients matrix.

We can first observe the considerable impact that both passive and active \gls{ris} can exert on the system performance, particularly in terms of \gls{ee}. This emphasizes the critical importance of optimizing both precoding and the matrix of reflecting coefficients of the \gls{ris} within the \gls{ris}-aided \gls{m-mimo} scenario since it can deal with the interference, improving the system \gls{se}, and consequently the system \gls{ee}. Furthermore, delving deeper into our findings, the result reveals that the proposed active \gls{ris} optimization method significantly improves the \gls{ee} of the \gls{ris}-assisted \gls{m-mimo} system compared to the entirely passive \gls{ris} configuration. Specifically, our result indicates that the active \gls{ris} can outperform
the performance of its entirely passive counterpart, achieving an impressive 
improvement of up to 120$\%$ in the peak value at $\approx 36$ dBm. This emphasizes the efficacy of deploying active \gls{ris} in maximizing the \gls{ee}, demonstrating its ability to deal with the double-fading attenuation drawback significantly.

\subsection{Percentage of Amplification Power Utilized}

\begin{figure}[!htbp]
\centering
\includegraphics[width=.69\columnwidth]{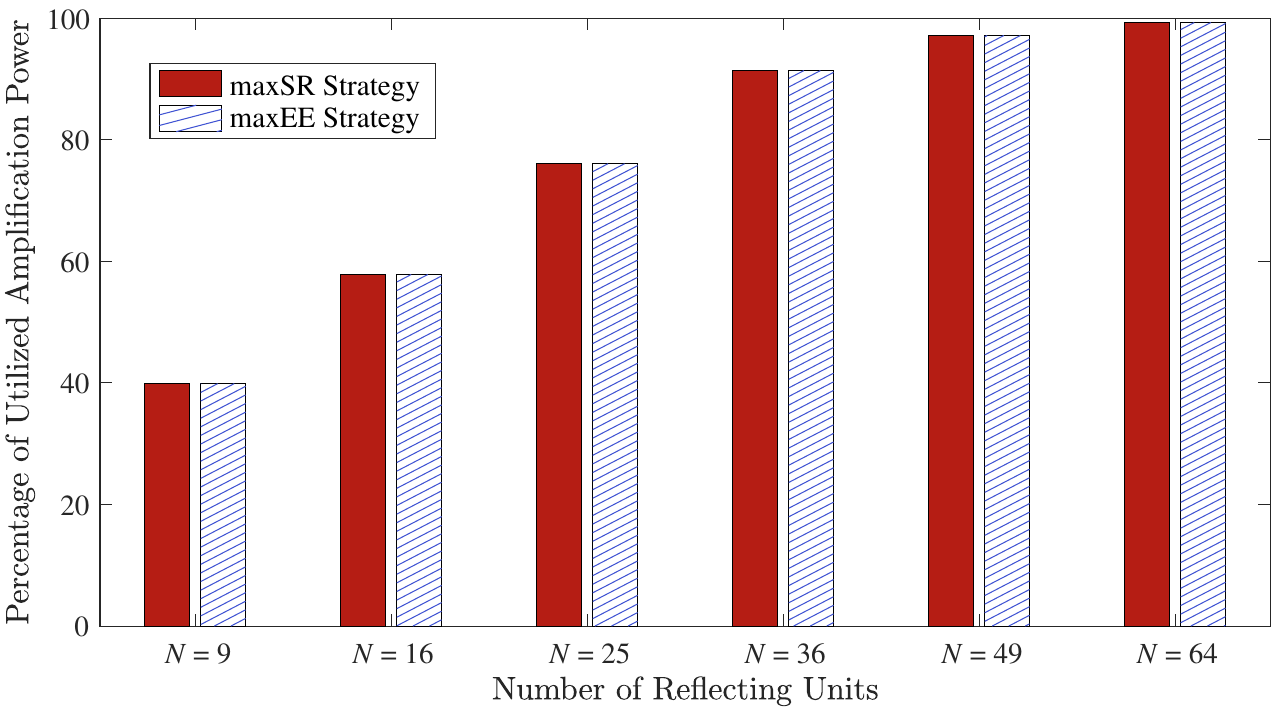}
\caption{Percentage of power amplification utilized {\it vs} the number of reflecting elements for the proposed algorithm. In this setup, we set $P_{\rm TX} = 35$ dBm.}
\label{fig:power}
\end{figure}

Figure \ref{fig:power} illustrates the percentage of utilized amplification power as a function of the number of reflecting elements for two distinct strategies: 1) \gls{sr} maximization, and 2) maximization of \gls{ee}. One can see through this result that amplification power $P_{\max}^{\rm RIS}$ increases as the number of reflecting elements increases. A reason for such a behavior is that when $N$ is low, utilizing fully $P_{\max}^{\rm RIS}$ can result in higher interference, and consequently poor \gls{ee} since the denominator of \eqref{eq:pEE} also increases with $P_{\max}^{\rm RIS}$. Besides, by increasing the number of reflecting elements, the \gls{ris} can easily achieve equality on the right side of constraint \eqref{eq:c1}, showing that as higher is $N$, better the \gls{ris} can manage the interference. Furthermore, one can see that the utilized power amplification for the \gls{sr} maximization strategy is also limited for scenarios when $N$ assumes low values, confirming that this power management is necessary for the optimization of both metrics.

\begin{figure}[!htbp]
\centering
\includegraphics[width=.69\columnwidth]{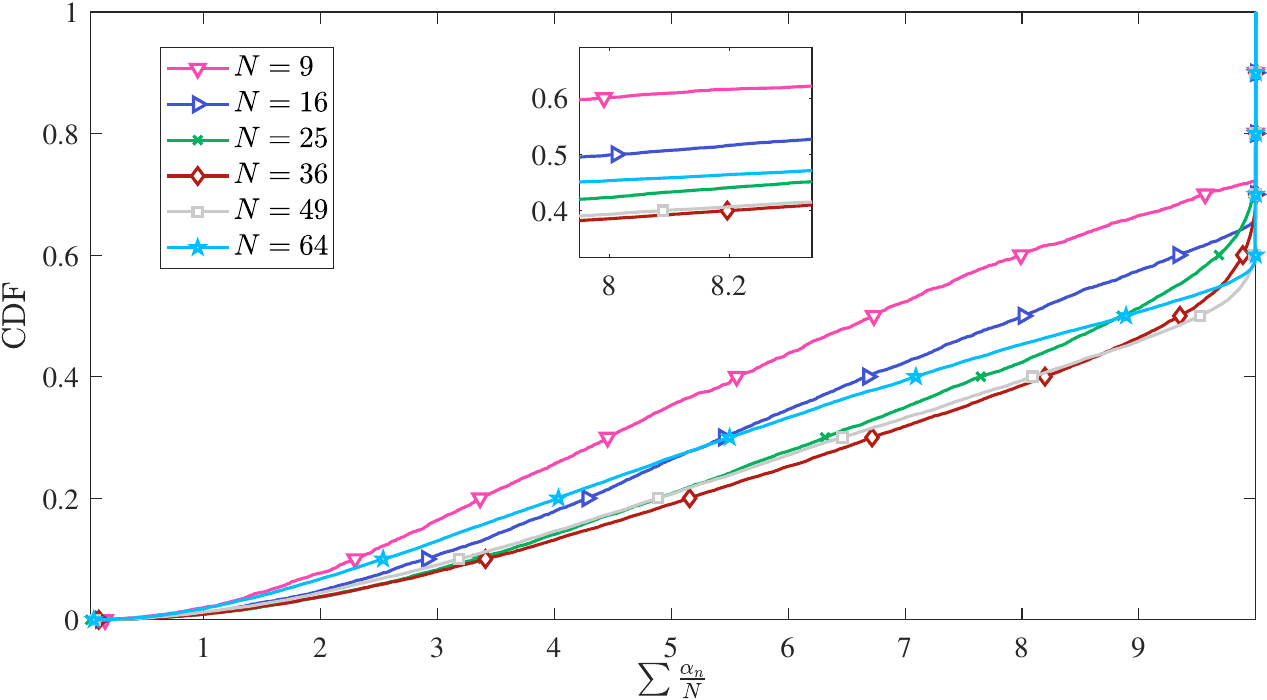}
\caption{CDF of the average value of amplitude obtained with the proposed algorithm. Here we set $P_{\rm TX} = 35$ dBm.}
\label{fig:alpha}
\end{figure}

In Figure \ref{fig:alpha}, we plot the \gls{cdf} of the {\it average amplitude gain} of the active \gls{ris}, $\bar{\alpha} =\frac{1}{N} \sum_{n=1}^N \alpha_n$, for some values of $N \in [9; 64]$ \gls{ris} elements. One can observe that the probability of $\bar{\alpha}$ to assume a low value, is higher for $N=9$, than $N=[16,25,36,49,64]$. This coincides with the result shown in Fig. \ref{fig:power}, justifying the fact that low $N$ values ($N=\{9,16\}$) utilize lower amplification power. However, for $N=64$, the algorithm also applies low values of $\bar{\alpha}$.
This is because, within scenarios where $N$ assumes high values, the maximum available amplification power is easily achieved. We can see that by increasing $N$ from $N=49$, the $\bar{\alpha}$ value starts to decrease.

\subsection{Number of RIS Reflecting Elements}

\begin{figure}[!htbp]
\centering
\includegraphics[width=.69\columnwidth]{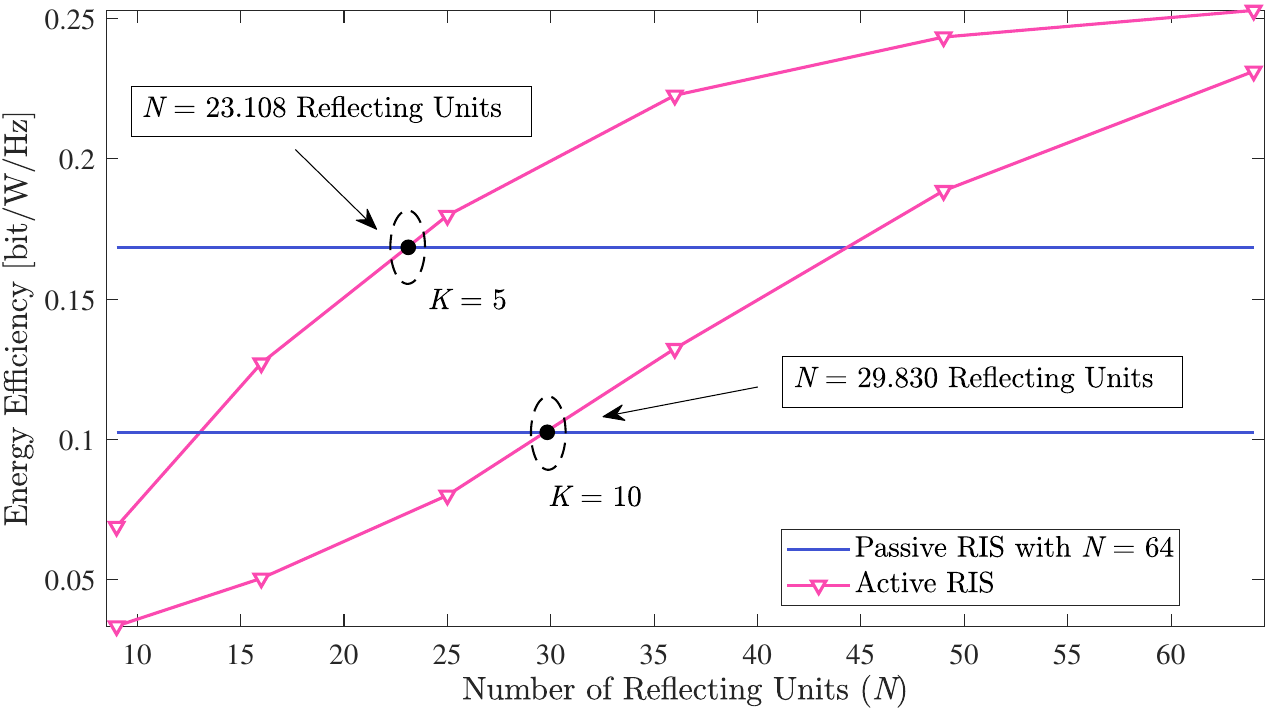}
\caption{Average \gls{ee} {\it vs} number of RIS reflecting units. Performance evaluation of the proposed algorithm for the active \gls{ris} {\it vs} entirely passive \gls{ris} with fixed number of elements $N=64$, for $P_{\rm TX} = 35$ dBm and two different number of \glspl{ue}, $K=5$ and $K=10$.}\label{fig:nbrN}
\end{figure}

In the sequel, Fig. \ref{fig:nbrN} illustrates the system \gls{ee} as a function of 
the number of reflecting elements ($N$) of \gls{ris}. Additionally, the performance of an entirely passive \gls{ris}, with a fixed number of elements $N=64$, is depicted in the plot. We can see that the active \gls{ris} exhibits comparable performance to the passive \gls{ris} when equipped with $\approx 23$ reflecting elements. In other words, an active \gls{ris} with $N = 25$ ($N_v = N_h = 5$) can surpass, in terms of \gls{ee}, the entirely passive \gls{ris}-aided M-MIMO with $N=64$ elements. Furthermore, we can see when we increase the number of \glspl{ue} $K$, the number of \gls{ris} elements is slightly affected. This again demonstrates the potential of amplifying the incoming signal at the \gls{ris}, enabling the possibility of reducing the number of elements on the \gls{ris} without compromising the system \gls{ee} performance. Additionally, it shows the potential of active \gls{ris} to reduce the channel estimation overhead, keeping the same performance as entirely passive \gls{ris}, which can be very useful for practical systems since the channel estimation overhead is one of the most critical drawbacks associated with the \gls{ris} technology.

This unveils an interesting aspect of the active \gls{ris}, that enables it to achieve the same \gls{ee} performance as passive \gls{ris} while reducing the size of the array physically. Let's assume elements with a size of $d = \lambda$ and an adjacent spacing between elements of $d_{\rm R} = \frac{\lambda}{2}$. For a frequency 
operation of $f = 5$ GHz, where the wavelength is $\lambda = 60$ mm, the length of \gls{ris} can be calculated using the formula $\mathcal{L} = \lambda N_h + \frac{\lambda (N_h - 1)}{2}$. With $N = 64$, the length is $\mathcal{L}=69$ cm, while for active \gls{ris} with $N=25$ the length is 42 cm. This demonstrates that the active \gls{ris} can achieve comparable performance with a physically smaller array, providing potential 
deployment advantages.

Moreover, envisioning a scenario where system priorities may shift between \gls{se} or \gls{ee} metrics, a dynamic approach to controlling the number of active \gls{ris} elements becomes 
paramount. For instance, when the \gls{se} is the priority, employing a higher quantity of operating elements is a natural choice.  However, when prioritizing \gls{ee}, a conscious reduction in the number of acting elements becomes crucial. This flexibility allows an opportunistic and consistent selection of RIS elements conditioned to the system's specific needs in different operational contexts.

\subsection{Convergence Performance of Proposed Algorithm}

\begin{figure}[!htbp]
\centering
\includegraphics[width=.69\columnwidth]{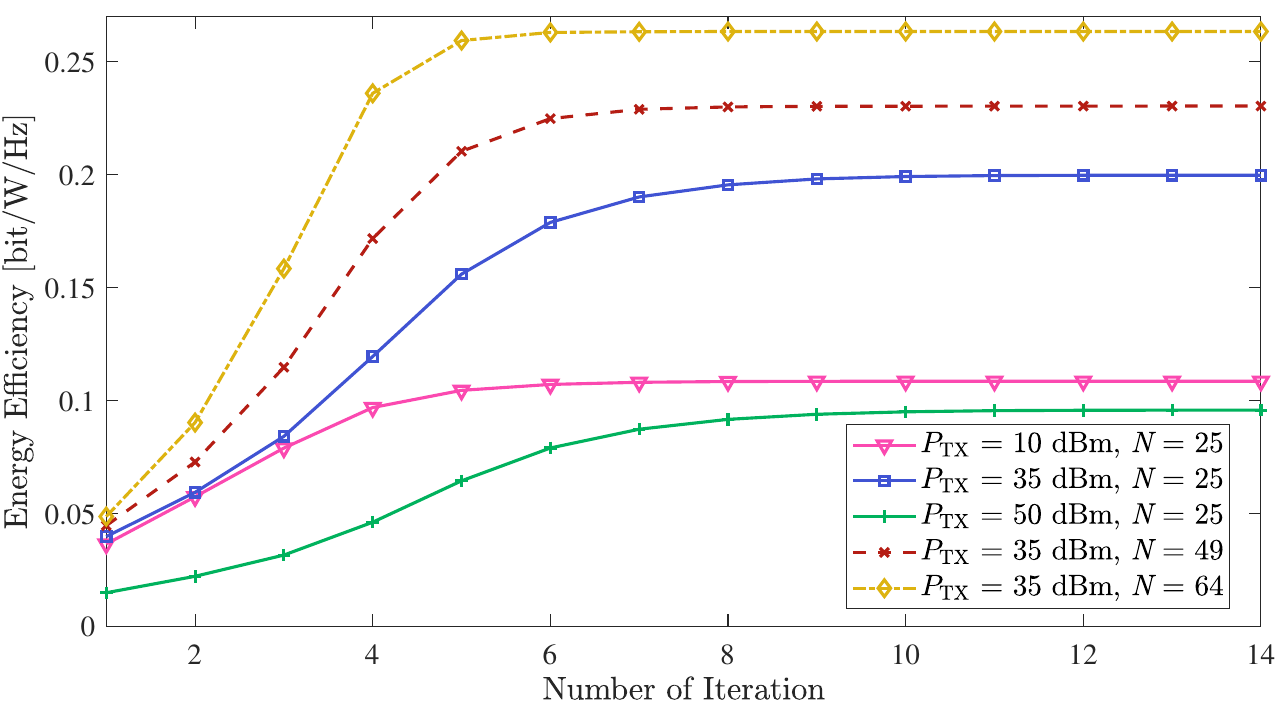}
\caption{Average \gls{ee} {\it vs} number of iterations for the proposed algorithm. Performance evaluation of the proposed algorithm for the active \gls{ris} with fixed number of elements $N=25$, for $P_{\rm TX} = \{10, 35, 50\}$ dBm and for fixed transmit power budget $P_{\rm TX} = 35$ dBm and $N = \{49, 64\}$.}
\label{fig:nbrI}
\end{figure}

Fig. \ref{fig:nbrI} illustrates the convergence behavior of the proposed algorithm under various conditions, showcasing their effectiveness for each one. The monotonic convergence patterns are examined across three distinct values of $P_{\rm TX}$ with constant $N$, and for the same $P_{\rm TX}$ under two different values of $N$. It is clear that the \gls{ee} initially experiences an upward trend and subsequently stabilizes with increasing iterations across all modes, highlighting the efficiency of the proposed algorithm. In addition, a noteworthy trend can be observed: as the number of reflective elements $N$ increases, so does the achieved \gls{ee}. Moreover, we can see that the proposed algorithm demonstrates rapid convergence (number of iterations) since it does not need many iterations (6 to 9 iterations) to converge, $e.g.$, for the case when $P_{\rm TX} = 35$, and $N=64$, the number of iteration necessary to converge is 6, while for $P_{\rm TX} = 50$, and $N=25$, the number of iteration necessary to converge is 9. This result shows that the power level has a more significant impact on convergence than the number of reflective elements, providing valuable insights into the algorithm's efficiency under varying operational conditions. 

\section{Conclusions
} \label{sec:concl}

We have examined the profound implications of 
optimizing the \gls{ee} for active \gls{ris} scenarios, seeking to understand and find the number of elements that need to be selected for acting in active \gls{ris} given outperforming its entirely passive \gls{ris} counterpart. By exploring various optimization techniques, we aim to provide a low-complexity solution and ensure its proximity to the optimal solution. This approach allows us to assess the effectiveness of our analytical methodology to handle complex optimization challenges. 

The numerical results show that the active \gls{ris} can easily outperform the entirely passive \gls{ris} in terms of system \gls{ee}, achieving a gain of about 120$\%$. Furthermore, for typical system and channel scenarios, active \gls{ris} can operate with less than half of elements compared with the passive \gls{ris} in view of achieving about the same performance, enabling an operational cost reduction for physical arrays and lower channel estimation overhead for practical implementations. However, to achieve better performance in signal amplification, the active \gls{ris} needs to be equipped with a reasonable number of reflective elements, mainly in scenarios where the available power for signal amplification is high. Finally, we show how the proposed algorithm can be promising since it has required a few iterations for convergence.

We delved into the optimization of energy efficiency (\gls{ee}) in \gls{ris}-assisted \gls{m-mimo} systems. While significant progress has been made, several crucial aspects pose open research challenges, offering promising avenues for exploration.
Our initial focus was on optimizing the \gls{ris} reflection coefficients within the framework of \gls{mr} precoding with \gls{epa}. However, it is noteworthy that the precoding strategy at the \gls{bs} also presents an opportunity for optimization. Considering the joint optimization of \gls{ris} and \gls{bs} precoding can potentially unveil enhanced \gls{ee} performance.

Moreover, our system model has assumed a far-field propagation channel. Extending and investigating the analysis of near-field propagation channels is a very relevant topic for future works. This extension is particularly pertinent due to the prevalence of large aperture arrays aimed at mitigating the multiplicative path-loss effect, and higher frequency operations for increased bandwidth, which inherently 
increase the Rayleigh distance. Notably, this has extended the range where near-field propagation dominates. Furthermore, new opportunities arise in the near-field context; for instance, the beam-focusing effect becomes a prominent phenomenon. This effect not only allows adjustment of the direction of the reflected beam but also offers the intriguing capability to control the distance within a specific angle. Developing strategies to improve the \gls{ee} of \gls{ris}-assisted \gls{m-mimo} networks leveraging the beam-focusing effect for near-field scenarios is thus of significant relevance.

Additionally, the current investigation centered on a single active \gls{ris}. Extending the system model to accommodate a scenario with multiple active \gls{ris} units is a valuable avenue for future research, offering insights into the complexities and opportunities associated with such configurations.

Finally, the optimization scope can be broadened by considering the joint optimization of \gls{bs} and \gls{ris} antenna elements. By optimizing both components simultaneously, the overall \gls{ee} of the system can be finely tuned for improved performance and resource utilization. These avenues highlight the richness of potential research directions in advancing the understanding and optimization of active \gls{ris}-assisted \gls{m-mimo} systems. 


\end{document}